\documentclass[12pt,titlepage]{article}
\usepackage{amssymb}
\usepackage{epsfig}
\renewcommand{\theequation}{\arabic{section}.\arabic{equation}}
\usepackage{bm}
\font\grande=cmr10 scaled \magstep4
\font\medio=cmr10 scaled \magstep2
\outer\def\beginsection#1\par{\medbreak\bigskip
      \message{#1}\leftline{\bf#1}\nobreak\medskip
\vskip-\parskip
      \noindent}

\newcommand{\eq}{\begin{equation}}

\newcommand{\eqx}{\end{equation}}
\newcommand{\eqn}{\begin{eqnarray}}

\newcommand{\bi}{\begin{itemize}}

\newcommand{\eqnx}{\end{eqnarray}}
\newcommand{\ei}{\end{itemize}}

\newcommand{\nn}{\nonumber}
\newcommand{\ra}{\rangle}
\newcommand{\la}{\langle}
\newcommand{\ls}{\lambda_s}
\newcommand{\lp}{\lambda_P}
\newcommand{\kt}{\boldsymbol{k}}
\newcommand{\pt}{\boldsymbol{p}}

\newcommand{\qt}{\boldsymbol{q}}

\newcommand{\bt}{\boldsymbol{b}}
\newcommand{\xt}{\boldsymbol{x}}

\newcommand{\eT}{\epsilon_T}
\newcommand{\eL}{\epsilon_L}
\newcommand{\Dr}{\frac{\partial}{\partial r^2}}
\newcommand{\A}{\mathcal{A}}
\newcommand{\Hc}{\mathcal{H}}
\newcommand{\Lc}{\mathcal{L}}

\begin{document}
\begin{flushright}
CERN-PH-TH/2007-254\\
DFF 440/12/07
\end{flushright}
\vspace{6mm}
\begin{center}

\grande{Towards an  S-matrix Description \\ of Gravitational Collapse}

\vskip 5mm
\large{ D. Amati}
\vspace{3mm}

{\sl SISSA, 34014 Trieste and INFN, Sezione di Trieste} 

\vspace{6mm}

\large{ M. Ciafaloni}

\vspace{3mm}

{\sl Dipartimento di Fisica, Universit\' a di Firenze}

{\sl  and}

{\sl INFN, Sezione di Firenze, 50019 Sesto Fiorentino, Italy} 

\vspace{6mm}

\large{ G. Veneziano}
\vspace{3mm}

{\sl Theory Division, CERN, CH-1211 Geneva 23, Switzerland} 

{\sl and} 

{\sl Coll\`ege de France, 11 place M. Berthelot, 75005 Paris, France} 

\vspace{6mm}

\centerline{\medio  Abstract}
\end{center}
{Extending our previous results on trans-Planckian  ($Gs \gg \hbar$)  scattering of light particles in quantum string-gravity we present a calculation of the corresponding $S$-matrix from the  region of large impact parameters   ($b \gg G\sqrt{s}>\ls$)  down to the  regime where  classical gravitational collapse is expected to occur.  
By solving the semiclassical equations of a previously introduced effective-action approximation, we  find that the perturbative expansion around the leading eikonal result diverges at a critical value $b = b_c = O(G\sqrt{s})$, signalling the onset of a new (black-hole related?) regime.  We then discuss the main features of our explicitly unitary $S$-matrix -- and of the associated effective metric -- down to (and in the vicinity of) $b = b_c$, and present some ideas and results on its extension all the way to the $ b \rightarrow 0$ region.
We find that  for $b<b_c$ the physical field solutions are complex-valued and the S-matrix shows additional absorption, related to a new production mechanism. The field solutions themselves are, surprisingly, everywhere regular, suggesting a quantum-tunneling -- rather than a singular-geometry -- situation.}

\vspace{5mm}

\vfill

\section{Introduction}
\setcounter{equation}{0}
In this paper we will discuss, within a quantum  string-gravity framework, the high-energy scattering of light particles  in a variety of  kinematical regimes.  We will resume, for this task, our twenty-years old  trans-Planckian $S$-matrix analysis ~\cite{ACV1, ACV4, ACV2, ACV3} and  we will extend it to the situation in which, at a classical level, the initial state is doomed to collapse due to the appearance of a closed trapped surface \cite{EG, KV}.

Renewed  interest in this problem stems from a growing conviction that a consistent quantum calculation of a collapse process leading to black-hole formation and to its subsequent evaporation is the best --if not the only-- way to understand the fate of the apparent information paradox \cite{IP}; or, better, the apparent loss of quantum coherence implied by black holes. Hybrid quantum arguments in terms of classical gravitational solutions do not meet, in our opinion, the necessary consistency requirements. It has indeed been suggested~\cite{Amati} that pure quantum states would not produce gravitational collapse even if the energy distribution would classically predict it to happen. And, even more drastically, it has been proposed that quantum back-reaction on the metric in apparently collapse-prone processes, would generate everywhere regular
solutions without singularities and event horizons ~\cite{QBR}. Finally, topologically non-trivial (i.e collapse-like)  classical  configurations may turn out to be  irrelevant in the quantum formulation of the physical process~\cite{Hawking2}.

Hints of what may actually happen has to come from the actual treatment of a collapsing system in a consistent quantum theory of gravity. Unfortunately, there are not many candidates for such a theory. To our point of view, among them, string theory is the  only  one allowing for a treatment of  the problem in perturbative, as well is non-perturbative regimes, despite the fact (or perhaps, because of the fact) that it is not, to start with, a general relativistic theory  describing space-time dynamics. Strings can only be consistently quantized in appropriate backgrounds, those that do not introduce two-dimensional Weyl anomalies. As in our previous papers, we will study the scattering process in $D=10$ superstring theory in Minkowski space-time -after compactifying $n$ dimensions on string-size tori- and look at possible (perhaps even approximate) interpretations of the results in terms of an {\it effective} metric  best describing the quantum process.

Let us briefly recall our (ACV hereafter) approach  and results.  Scattering of two massless strings (e.g. of two gravitons) was considered at centre of mass energy $2 E=\sqrt s \gg M_{\rm {Planck}}$ and impact parameter $b$ in $d$-dimensional Minkowski  spacetime, where $d= 10-n$. In this paper we shall work in $d=4$, but we expect that the extension to $d>4$ will not present major problems. We shall also focus on a regime in which string-size effects  are relatively small, while the gravitational interactions can be  strong. In order to define this more precisely, let us recall that, in string-gravity, the fundamental scale is the string length $\ls=\sqrt{\alpha'\hbar}$, in terms of which the Planck length and the Newton constant are expressed as $\lp=\sqrt{G\hbar}=g\ls$, where $g\ll 1$ is the string loop expansion parameter, assumed to be small.

On the other hand, at very high energies $\sqrt{s}$, the gravitational (Schwarzschild) radius $R=4GE=2G\sqrt{s}$ plays an important role. In our small-coupling, high-energy regime, defined by $Gs>>\hbar$, $R$ is much larger than $\lp$, but can be smaller or larger than $\ls$, because the ratio $R/\ls=g\sqrt{Gs/\hbar}$ involves the small coupling constant $g$.  As a consequence, there are three distinct regimes according to which one of the three length scales $b$, $R$, or $\ls$ exceeds the other two.

If $b \gg R, \ls$ one deals with small deflection-angle scattering. This is well decribed by a leading eikonal approximation with small string-size and classical corrections corresponding to the expansion parameters $(\ls/b)^2$  and $(R/b)^2$, respectively. The former are quite easily taken into account \cite{ACV1} and can best be interpreted \cite{SSE} as string excitations due to  the tidal forces induced on each string by the effective (Aichelburgh-Sexl) shock-wave metric produced by the other string. They have been analyzed by ourselves in the past \cite{ACV1}  and, most recently, in \cite{GGM}

In the regime $\ls \gg  b, R$, also investigated through fixed-angle scattering \cite{Gross}, string effects soften gravity according to the generalized uncertainty relation~\cite{EUP, ACV4}
\eq\label{UR}
\Delta x > \frac{\hbar}{\Delta p} + \alpha' \Delta p > \ls.
\eqx
As a consequence, the minimal observable size of the system is $\ls$ itself, which exceeds $R$, and  classical gravitational collapse conditions are  never met. It is possible, however, to push the analysis of this regime towards its  boundary $\ls \rightarrow  R >b$,  which should correspond to the threshold for black hole formation $E_{\rm{threshold}} \sim M_s g^{-2} \sim M_P g^{-1}$. One finds \cite{GV, ACV1} that, even if no black-hole is formed, the final state, in the energy region $M_P < E < E_{\rm{threshold}}$  starts to vaguely resemble that of an evaporating black hole of mass $\sqrt{s}$ with typical final momenta of order $M_P^2/\sqrt{s}\simeq\hbar/R$. In other words, a precocious black-hole-like behaviour is found to occur even below the expected threshold for their actual production.

By contrast, for $R>\ls$, new semiclassical phenomena take place. They  extend beyond the impact parameter at which string fluctuations, including those due to diffractive excitations, are large. This is the regime that we attempt to treat in this paper, for various values of the impact parameter $b > \ls$ of the colliding strings. The interesting region is the one in which $b$ approaches $R$ from above and possibly goes below it, a situation in which, classically, a gravitational collapse would take place \cite{EG}. A general framework for describing this most difficult regime was proposed in \cite{FPVV}, where the S-matrix was connected to properties of a classical solution at past and future null infinity (the so-called Bondi masses). However, in spite of its conceptual appeal, going beyond the leading eikonal in that formalism has proven prohibitively difficult. Here we shall use instead our key observation \cite{ACV1}  that, because of the softness of multi-loop string amplitudes, the S-matrix exponentiates in terms of an eikonal function of order $Gs/\hbar$ which, in turn,  can be expanded in powers of $R^2/b^2$ for $b>R$. The outcome has a diagrammatic interpretation that can be encoded into an effective action. If string effects are neglected, that action agrees with Lipatov's effective action~\cite{Lip} (see also \cite{verlinde2}) and reproduces~\cite{ACV3} the previously computed leading-order correction to the eikonal ~\cite{ACV2}.

Therefore, the effective Lagrangian that we investigate in this paper is motivated by our string-gravity expansion, even if it does not contain explicit string corrections. It is a function of appropriate components of the metric  $h_{\mu\nu}(x)$ which are apt to describe the high-energy regime, interact among themselves via the effective coupling $R/b$, and are coupled to sources provided by the energetic scattering particles. The solutions of the (nonlinear) lagrangian equations provide an effective metric (which appears as the outcome of quantum backreaction effects) in terms of which the action, and thus the S-matrix, is expressed and computed. The unitarity of the approach implies an absorption in the elastic amplitude due to particle production, whose inclusive properties (spectra, correlations, etc...) may be analyzed.

Within this framework, we treat here the region $b\gtrsim R$ and we attempt to tackle the most interesting region $b\lesssim R$ where still one should be able to compute the scattering amplitudes and the effective metric. We cannot claim to have fully achieved that goal, but we have progressed pretty far towards it, even beyond expectations. We thus believe that the analysis of our results should provide at least some hints as to whether - in this consistent quantum approach - there is any sign of a trapped region or event horizon and what is the ``unitary evaporation'' that is produced without loss of quantum coherence. 

The rest of the paper is organized as follows:

In Section 2 we recall the eikonal expansion,  the form of the first correction to its leading term, and the effective action that should generate through its tree diagrams the higher  order corrections. We then define a somewhat simpler problem in which a ``rescattering'' term, the related ``double-diffractive'' string excitation, as well as one of the emitted-graviton polarizations, are neglected. The above approximations will be used throughout the paper, although, in Section 6, we will give an educated guess on how the second (infrared-sensitive) polarization can be included.

In Section 3 we discuss the axisymmetric case in which the field equations become ordinary differential equations. We first consider  a class of analytical solutions to the field equations in the  case of point-particle collisions at $b=0$. We are aware of the fact that this is a most difficult regime for justifying some of our approximations, in particular the neglect of string corrections, which should be restored later on. Nevertheless, this class of $b=0$ solutions -- which is surprisingly simple and robust, but quite non-perturbative -- turns out to tune up the discussion on the boundary conditions to be set in order to match perturbation theory at larger values of $b\gtrsim R$. Furthermore, they are likely to be essential for the overall interpretation of the problem for $\ls < b\ll R$.  As an amusing digression we will also discuss here the central collision of two extended sources (taken to be two identical homogeneous disks of radius $\Sigma$ for simplicity) where the problem can be solved analytically and shows the existence of a critical ratio $\Sigma/R$. This case could also be a way to represent string-string collisions with $\Sigma \sim \lambda_s$.

In Section 4 we turn to the case of generic values of $b$ where, within some technical approximations, we are still able to solve the problem analytically.  We find that, while at $b \gg R$ the perturbative expansion is qualitatively correct,  the expansion diverges at some calculable critical value of $b = b_c  \sim R$. We also discuss possible ways to define solutions below $b=b_c$.

In Section 5 we reformulate the problem in momentum space as a set of integral equations lending themselves to an iterative solution.  We reach conclusions that are in very good agreement with those obtained in the position space approach of the previous section: in particular, the iterative solution only converges above a critical value of $b/R$.

In Section 6 we turn our attention to the construction of a unitary $S$-matrix,  to the properties of the final state and to the expectation value of the metric in that state. Once more, the properties of the final state appear to resemble those due to an evaporating black hole as we approach $b_c$.

In Section 7 we describe (and try to interpret) our proposal to define the scattering amplitude and effective metric for $b<b_c$, which is based on the analysis of properly identified complex-valued solutions of the field equations. Finally, in Section 8, we  summarize  our main results and give a brief outlook.

\section{Eikonal Expansion and effective action}\label{s:eexp}
\setcounter{equation}{0}

\begin{figure}[htbp!]
\centering{
  \includegraphics[width=0.65\textwidth]{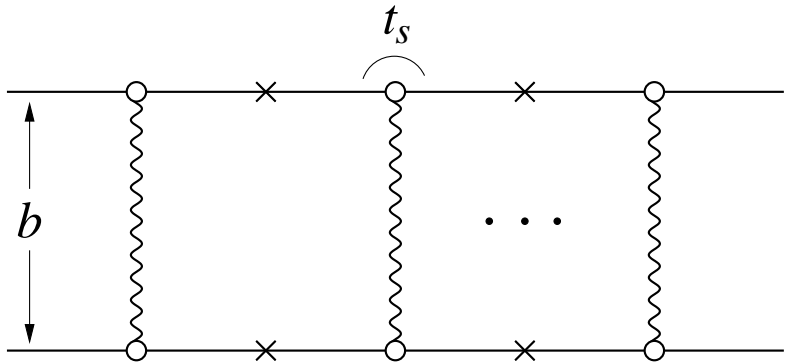}\\}
  \caption{\it The leading eikonal rescattering series. Crosses denote on-shell propagation
    \label{1}}
\end{figure}

In string-gravity, ACV found that the S-matrix in the impact parameter representation has an eikonal form where the eikonal operator can be expanded in the parameter $R^2/b^2$. For $b>>R>\ls$, the eikonal resums all powers of $Gs$ due to multigraviton exchanges, as follows (see also \cite{others}):
\eq\label{eikonal1}
S(b,s) = \exp{2i\delta_0(b,s)},~~~ \delta_0=\frac{Gs}{\hbar} \log(\frac{L}{b})\, ,
\eqx
where $L$ is an infrared cutoff related to the well known infinite Coulomb phase. String effects in this region are simply taken into account~\cite{ACV1} by an operator shift of the impact parameter variable
\eq\label{eikonal2}
\delta_0 \rightarrow \hat{\delta}_0 = \delta_0 (b+\hat{X}_u-\hat{X}_d,s)\, ,
\eqx 
(where an average is performed over the closed string position operators $\hat{X}_u, \hat{X}_d$)
and give rise to the diffractive string excitation and fluctuations mentioned before. The soft behavior of multi-loop string amplitudes is itself responsible for the dominance of eikonal iteration in the results (\ref{eikonal1}), (\ref{eikonal2}). In fact, Eq.(\ref{eikonal1}) can be interpreted (fig.~(\ref{1})) as a multiple scattering series, in which the (small) deflection angle $\theta=2R/b$ - corresponding to a possibly large momentum transfer $t\simeq Gs\hbar/b^2$ - is built up by many graviton-exchange processes of small momentum transfer, of order  $t_s \simeq (\hbar/b)^2 \ll (\hbar/\ls)^2$.

\begin{figure}[htbp!]
\centering{
  \includegraphics[width=0.6\textwidth]{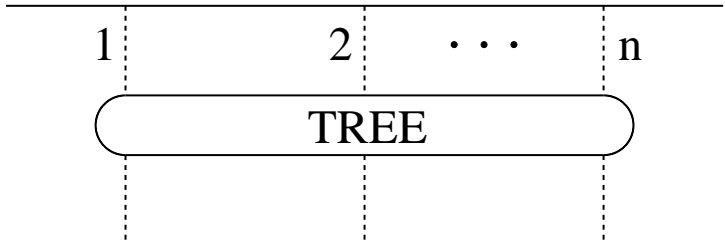}\\}
  \caption{\it Diagrammatic structure of $2n$-loop irreducible contributions to the eikonal  
    \label{2}}
\end{figure}

When $R/b$ becomes sizeable, ACV found that the eikonal can be expanded in a power series in $R^2/b^2$ and, possibly, $\ls^2/b^2$. The terms $(R^2/b^2)^n$ of such a series are in correspondence with connected tree diagrams interacting with the colliding strings via the exchange of $2n$ (reggeized) gravitons, as shown in fig.~(\ref{2}). Besides the one-loop correction $\delta_1(b,s)$, ACV found that the lowest term in such a series is the so-called H-diagram of fig.~(\ref{3}), contributing at two loop level to the real part of $\delta_2$
\eq\label{delta2}
\delta_1(b,s)~=~\frac{Gs}{\hbar}~\frac{6\ls^2}{\pi b^2};~~~\mathrm{Re}\delta_2(b,s)~=~\frac{Gs}{\hbar}~\frac{R^2}{2b^2}\, .
\eqx
This extra contribution to the phaseshift modifies the Einstein deflection angle of energetic (massless) particles in the form:
\eq\label{theta2}
\sin\frac{\theta_{cl}}{2}~=~\frac{R}{b}(1+\frac{R^2}{b^2}+...) \, .
\eqx
Further terms in the expansion are expected, and will be calculated here in a framework to be defined shortly.

ACV found also inelastic effects contributing to the imaginary part of the phaseshift. There is an infrared divergent contribution, connected with soft graviton bremsstrahlung, not explicitly discussed here, and a finite part $\mathrm{Im}\delta_2=2\log s~\mathrm{Re}\delta_2/\pi$, connected with hard graviton emission, which will be generalized in the following.

\subsection{H-diagram: Amplitude and emission field} \label{s:Hdiag}
\begin{figure}[htbp!]
\centering{
  \includegraphics[width=0.45\textwidth]{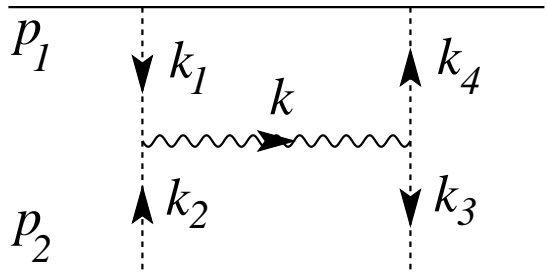}\\}
  \caption{\it Regge-Gribov H-diagram, yielding the first subleading correction to the eikonal; dashed (wavy) lines denote exchanged (emitted) gravitons corresponding to the fields $a_0$ and $\bar{a}_0$ ($h_0$ or $\phi_0$)
    \label{3}}
\end{figure}

In order to extract from the H-diagram in fig.~\ref{3} a computational method for higher orders in $R^2/b^2$ of the eikonal operator, we have introduced in ref.~\cite{ACV3} an effective action approach. An essential ingredient in it are the high-energy graviton~\cite{Lip2} and string~\cite{ABC} emission vertices, which lead to the emission amplitude that we now recall. 

While the leading eikonal exponential is generated by the exchange among the colliding particles' sources of an arbitrary number of (longitudinal) gravitons, the $\sim R^2/b^2$ correction is represented by a graviton being emitted by two exchanged ones and then absorbed by two others. We shall see that this intermediate graviton, in its  transverse polarization,  will play an important role in the scattering process. To this purpose, let us define the following two independent transverse-traceless polarization tensors for a graviton of momentum $k$ ( bold-case notation referring to transverse momenta):
\eqn\label{pol_a}
\epsilon^{\mu\nu}_{TT} &=&  (\eT^{\mu}\eT^{\nu}-\eL^{\mu}\eL^{\nu}) \, ,  \, ~\epsilon^{\mu\nu}_{LT} =  
(\eL^{\mu}\eT^{\nu}+\eT^{\mu}\eL^{\nu}) , ~~~\epsilon_{\mu\nu}^i\epsilon^{\mu\nu}_j=2\delta^i_j~~~(i,j=TT,LT)\, ,
\eqnx
where:
\eq\label{pol_b}
\eL^\mu+i\eT^\mu \equiv \eta^\mu(k)=(\frac{k^3}{|\kt|},i\boldsymbol {\epsilon},\frac{k^0}{|\kt|})\, ,
\eqx
and $\boldsymbol {\epsilon}$ is the unit polarization vector transverse to $\kt$. 

At high energies, on the basis of the vertices in~\cite{Lip2}, \cite{ABC}, the graviton emission amplitude of fig.~(\ref{3}) takes the form ($\hbar=1$)
\eqn\label{Hvertex}
A^{\mu\nu}&=&\frac{\kappa^3 s^2}{\kt_1^2 \kt_2^2\kt^2} \mathrm{Re}[(k_1^2{k^*_2}^2-|k_1|^2|k_2|^2)\eta^\mu\eta^\nu]\\  \nn
&=&\frac{2\kappa^3s^2}{\kt^2}[{\sin^2 \theta_{12}} \epsilon^{\mu\nu}_{TT} - \sin \theta_{12}\cos \theta_{12}\epsilon^{\mu\nu}_{LT}]\, ,
\eqnx
where we have defined $\kappa^2=8\pi G$ and $\kt=\kt_1+\kt_2= \kt_3 + \kt_4$.
From eq.~(\ref{Hvertex})  we derive the imaginary part of the H-diagram \cite{ACV2}
\eq\label{ImH}
\frac{\mathrm{Im}A_H (s, \qt^2)}{s}=\frac{Y}{16\pi s^2}\int d[\kt_1]d[\kt_2]A^{\mu\nu}(1,2)A^*_{\mu\nu}(3,4)
\, ,
\eqx
where $d[\kt]=d^2\kt/(2\pi)^2$, $Y=\log s$ and  $\qt=\kt_2-\kt_3$.  We
then obtain the real part by a dispersion relation which amounts to multiplication by $\pi/2Y$, and finally the impact parameter amplitude by a Fourier transform \footnote{We use conventions in which $4 s\delta(b,s) = \int d[\qt]~e^{- i\bt\qt} A(\qt, s)$ and ${\rm Im} A(0,s) = s \sigma_{tot}$.}:
\eqn\label{ReH}
\mathrm{Re}~\delta_H(b,s)=\frac{(8\pi G)^3s^2}{16}\int d[\kt_1]d[\kt_2]d[\qt]\frac{e^{- i\bt\qt}}{\kt^4}({\sin^2\theta_{12}} {\sin^2\theta_{34}} + \frac{\sin2\theta_{12}\sin2\theta_{34}}{4})\, .
\eqnx

By then introducing a $\kt_4$-integration through a delta-function $\delta(\kt_1+\kt_2-\kt_3-\kt_4)$ we can rewrite eq.~(\ref{ReH}) in two different convenient forms. The first, to be used in secs. (5) and (6),  uses the simple identity
$2 \qt=(\kt_2-\kt_1) + (\kt_4-\kt_3)$ to yield  the factorized form:
\eqn\label{FFmom}
\mathrm{Re}~\delta_H(b,s) = \frac{\pi}{2Y} \mathrm{Im}~\delta_H(b,s) = \frac{\pi}{2} Gs (\pi R)^2 \int d[\kt]\left( |h_{TT}(\kt)|^2 +  |h_{LT}(\kt)|^2 \right) \, ,
\eqnx 
where we have introduced the contributions of the $TT$ and $LT$ polarizations in eq.~(\ref{pol_a}):
\eqn\label{htthlt}
h_{TT}(\bt,\kt) &=& 16 \pi^2  \int \frac{d[\kt_1] d[\kt_2]}{\kt^2} \delta (\kt - \kt_1- \kt_2) 
\exp{\left(i \bt \kt_2)\right)}\sin^2\theta_{12}  \, , \\
h_{LT}(\bt, \kt) &=& 16 \pi^2 \int \frac{d[\kt_1] d[\kt_2]}{\kt^2} \delta (\kt - \kt_1- \kt_2) 
\exp{\left(i \bt \kt_2)\right)}\sin\theta_{12}\cos\theta_{12} \, .
\eqnx
Alternatively, we can rewrite eq.~(\ref{ReH}) as an $\xt$-space integral 
\eq\label{ReHx}
\mathrm{Re}~\delta_H(b,s)=\pi Gs\frac{(\pi R)^2}{2} \int d^2\xt (|h_{TT}(\bt,\xt)|^2+|h_{LT}(\bt,\xt)|^2)\, ,
\eqx
where  the analogous contributions of the $TT$ and $LT$ polarizations in positions space read:
\eqn\label{Hbypol}
h_{TT}(\bt,\xt)&=&4 \int \frac{d[\kt_1]d[\kt_2]}{(\kt_1+\kt_2)^2}\sin^2\theta_{12} \exp{i(\kt_2 \bt-\kt \xt)}
=\frac{1}{\pi^2}\frac{\sin^2\theta_{\bt\xt}}{|\bt-\xt|^2}\\ \nn
h_{LT}(\bt,\xt)&=&4 \int \frac{d[\kt_1]d[\kt_2]}{(\kt_1+\kt_2)^2}\sin\theta_{12}\cos\theta_{12}\exp{i(\kt_2 \bt-\kt \xt)}\, .
\eqnx 

Here we note that $h_{TT}$ in eq.~(\ref {Hbypol}) has a simple expression, which curiously reproduces the form of its Fourier transform and has a $1/\xt^2$ behaviour at large distances. On the other hand, $h_{LT}$ is more involved and shows a $1/|\xt|=1/r$ behaviour for $r\gg b$ that, in turn, produces the well-known logarithmic infrared divergence in eq~(\ref{ReH}), due to graviton bremsstrahlung. Indeed, one can show \cite{ACV2} that the $LT$ polarization is responsible for the Weinberg current~\cite{Weinberg}. The corresponding infrared behaviour was discussed in detail in ref.~\cite{ACV2}, where a subtraction in dimensional regularization was performed in order to obtain the finite result in eq. (\ref{delta2}) for $\delta_2$. On the other hand, here we are interested in the possibly collapsing energy, not in the one which is peripherally radiated away. Therefore, in most of the following, we will subtract the $LT$ polarization altogether, by restricting ourselves to the $TT$ one, which is IR safe.

Let us note some interesting properties of the $h$-fields just introduced. By defining $z=x^1+ix^2$ and $\partial=\partial/\partial z$, we find that the  complex combination $h_0=h_{TT}+ih_{LT}$, by the $i\sin\theta_{12}\exp(-i\theta_{12})$ form of the couplings in eq.~(\ref{Hbypol}), satisfies the differential equation
\eq\label{h_0}
2|\partial|^2 h_0(\bt,\xt)=4(\partial^2 a_0{\partial^*}^2 \bar{a}_0 - |\partial|^2a_0|\partial|^2\bar{a}_0)=\frac{1}{\pi^2}\frac{1}{{z^*}^2(b-z)^2}\, ,
\eqx
where we have defined the fields $a_0$ and $\bar{a}_0$ -- to be related to longitudinal gravitons -- by
\eq\label{a_0}
 a_0(z)=-\frac{1}{2\pi}\log(\frac{|z|^2}{L^2}),~~~ \bar{a}_0=a_0(b-z);~~~|\partial|^2a_0=-\frac{1}{2}\delta(\xt)\, .
\eqx
One may also double check that the expression for  $h_{TT}$ given in (\ref{Hbypol}) satisfies the real part of (\ref{h_0}) i.e.
\eq\label{Reh_0}
\nabla^2 {\rm Re}  h_0 = \nabla^2   h_{TT} = \frac{2}{\pi^2 \xt^2 (\bt - \xt)^2} \left( \frac{2 \left(\xt(\bt - \xt) \right)^2}{ \xt^2 (\bt - \xt)^2}  -1 \right)\, .
\eqx
Furthermore, one can define the Fourier transform of eq. (\ref{Hvertex}):
\eqn \label{FTA}
\tilde{A}^{\mu\nu} =  2 \kappa^3 s^2 \int \frac{d[\kt_1]d[\kt_2]}{(\kt_1+\kt_2)^2}\left( \sin^2\theta_{12} ~\epsilon_{TT}^{\mu \nu} - \frac12 \sin 2\theta_{12}~ \epsilon_{LT}^{\mu \nu} \right) \exp{i(\kt_2 \bt-\kt \xt)}\, ,
\eqnx
 and an effective gravitational field related to the H-diagram. This can be written in terms of the $h_0$ field as
\eq\label{metric_TT}
\frac{\tilde{A}^{ij}}{s}=\frac{\kappa^3s}{2}\tilde{h}_0^{ij}(\xt) = \frac{\kappa^3s}{2}\mathrm{Re}[\hat{\epsilon}^i\hat{\epsilon}^j h_0(\bt,\xt)],~~~ \tilde{h}_0^{ij}=\hat{\epsilon}^i\hat{\epsilon}^j h_0(\bt, \xt)= \frac{\delta_{ij} \nabla^2 - \partial_i\partial_j}{\nabla^2}h_0\, ,
\eqx
where we have promoted the polarizations $\epsilon^i$ to operators in $\xt$-space. The result is  better rewritten by introducing a (generally) complex scalar field $\phi$ such that \footnote{Here the field $\phi$ has a more convenient normalization, for  writing the action,  than that of ref.~\cite{ACV3}.}.
\eq
h=4|\partial|^2\phi=\nabla^2\phi;~~~~~\tilde{h}_{ij}= ( \delta_{ij} \nabla^2 - \partial_i\partial_j ) \phi\, .
\eqx 
In this language, restricting to the IR safe polarization means considering  $\mathrm{Re}\phi$ only, or the $\phi$ field to be  real. By replacing in eq.~(\ref{ReHx}) the corresponding expression of $h_{TT}$ in eq.~(\ref{Hbypol}), we obtain:
\eq
[\mathrm{Re}\delta_H(b,s)]_{TT}~=~\pi Gs \frac{(\pi R)^2}{2}\int d^2\xt (\nabla^2\mathrm{Re}\phi_0)^2~=~Gs\frac{3R^2}{8b^2} \, ,
\eqx
to be compared to the full result Re$\delta_2(b,s)=Gs \frac{R^2}{2b^2}$ of  eq.~(\ref{delta2}), the difference being due to the (neglected)  $LT$ polarization.

\subsection{The reduced effective action}\label{s:model}

\begin{figure}[htbp!]
\centering{
  \includegraphics[width=0.45\textwidth]{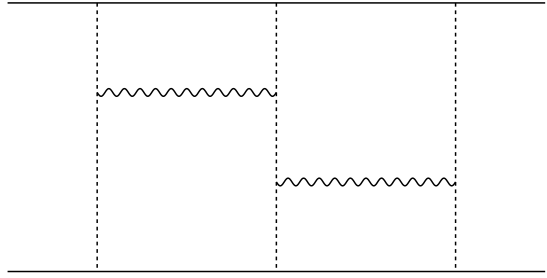}\\}
  \caption{\it Regge-Gribov double-H diagram, contributing at order $R^4/b^4$ to the eikonal; dashed (wavy) lines denote exchanged (emitted) gravitons 
    \label{4}}
\end{figure}

Further terms in the $R^2/b^2$-expansion are obtained by considering both multi-H diagrams combining multiple emissions (fig.~(\ref{4})) and rescattering diagrams in which emitted gravitons reinteract by a longitudinal exchange (fig.(~\ref{5})). Here we shall limit ourselves to the first class of diagrams, that will be treated to all orders, while the second class - which starts at order $R^4/b^4$ - will be briefly discussed in the next subsection.

Multi-H diagrams are described by a (reduced)  two-dimensional action introduced in ref.~\cite{ACV3} in which the longitudinal fields $a$ and $\bar{a}$ and the mostly transverse field $h$ occur, by generalizing the contributions of eqs.~(\ref{h_0}) and (\ref{a_0}). By restricting ourselves to the IR safe polarization, $h$ and $\phi$ are now supposed to be real valued and the reduced action takes the form
\eqn\label{reduced_a}
\frac{\A}{2\pi Gs}~&=&~a(\bt)+\bar{a}(0)-\frac{1}{2}\int d^2\xt \nabla \bar{a}\nabla a+\frac{(\pi R)^2}{2}\int d^2\xt(-(\nabla^2\phi)^2+2\Hc \nabla^2\phi) \nn \\ 
-\nabla^2 \Hc~&\equiv &~\nabla^2 a~\nabla^2\bar{a}-\nabla_i\nabla_j a~\nabla_i\nabla_j\bar{a}\, .
\eqnx
Here the longitudinal fields $a(\xt)$, $\bar{a}(\xt)$ interact with point-like sources placed at $\xt=b$  and $\xt=0$, respectively. The  field $\phi$ is generated by the current  $\Hc$, which is defined by a generalization of eq.~(\ref{h_0}) and expressed through vector derivatives, as it's more appropriate for real-valued fields. Note in (\ref{reduced_a}) the appearance of the effective coupling $R^2=4G^2s$, which controls  the dependence of the solutions on the expansion parameter $R^2/b^2$.

The coupled lagrangian equations derived from eq.~(\ref{reduced_a}) read
\eqn\label{reduced_eom}
\nabla^2 a + 2\delta(\xt)~&=&~2(\pi R)^2(\nabla^2a~\nabla^2\phi-\nabla_i\nabla_ja~\nabla_i\nabla_j\phi),~~~\bar{a}(\xt)=a(\bt-\xt) \nn \\ 
\nabla^2\Hc=\nabla^4\phi~&=&~-(\nabla^2a~\nabla^2\bar{a}-\nabla_i\nabla_ja~\nabla_i\nabla_j\bar{a})\, .
\eqnx 
It is soon apparent that $\Hc=h=\nabla^2\phi$ on the equations of motion, so that the ``on-shell'' action reads
\eqn\label{action_os}
\frac{\A(b,s)}{2\pi Gs}~&=&~a(b^2)+\bar{a}(0)-\frac{1}{2}\int d^2\xt \nabla \bar{a}\nabla a +\frac{(\pi R)^2}{2}\int d^2\xt(\nabla^2\phi)^2 \nn  \\ 
&=&~2a(\bt) + I_a + I_{\phi}\, .
\eqnx
 By performing the first perturbative iteration, we find thet $a$, $\bar{a}$ and $h$ reduce to the expressions in eqs.~(\ref{a_0}) and (\ref{h_0}), while the low order contributions to the action become
\eq
I_a^{(0)}=-a_0(\bt),~~I_a^{(1)}=-2a_1(\bt)=-4I_{\phi}^{(1)}\, ,
\eqx
so that, by collecting all terms, they partly cancel and finally yield
\eq\label{pert_ac}
\A(b,s)~=~2\pi Gs(a_0(\bt)+I_{\phi}^{(1)})~=~2\pi Gs~a_0(\bt)+2\mathrm{Re}~a_H~=~2Gs(\log\frac{L}{b}+\frac{3R^2}{8b^2})\, ,
\eqx
thus reproducing the perturbative result for the $TT$ part.

The framework defined by eqs.~(\ref{reduced_a}) and (\ref{reduced_eom}) is the one we shall analyze in detail in the following, including its features in the gravitational collapse region in which the effective coupling $R$ may exceed the physical size of the system $b$, larger than the string size $\ls$. Besides lacking the $LT$ polarization -- that we have argued to be related mostly to an infrared, peripheral phenomenon -- the above model lacks rescattering and string effects, whose form we shall briefly recall in the framework of an effective action \cite{Lip, verlinde2} which includes the dependence on the light-cone variables $x^{\pm}\equiv x^0\pm x^3$. This will allow us to relate the reduced action to a shock-wave solution  having a particular form in the $x^+, x^-$ plane.

\subsection{Shock-wave interpretation} \label{s:action}

The reduced action model introduced above can be viewed as a particular limit of Lipatov's effective action~\cite{Lip, verlinde2}, which in turn provides a formal description of the diagrammatic series in Fig.~(\ref{2}) in the limit in which all subenergies among emitted gravitons are large and string excitations are neglected. The corresponding Lagrangian contains, besides a field $\Phi$ (related to the previously introduced $\phi$), the longitudinal fields $\tilde{h}^{++}$ and $\tilde{h}^{--}$ which are similarly related to $a$ and $\bar{a}$ and are coupled to the external sources of the impinging particles (gravitons or strings)\footnote{As better explained in sec. 6, our fields are related to the usual metric components by: $ h_{\mu\nu}dx^{\mu}dx^{\nu}  \equiv ds^2- \eta_{\mu\nu} dx^\mu dx^\nu =2\kappa(\tilde{h}_{++}(dx^+)^2+\tilde{h}_{--}(dx^-)^2)+ (\kappa /4) (\epsilon_{\mu\nu}^{TT}\Delta\mathrm{Re}\Phi-\epsilon_{\mu\nu}^{LT}\Delta\mathrm{Im}\Phi) dx^{\mu}dx^{\nu} $}. 

\begin{figure}[htbp!]
\centering{
  \includegraphics[width=0.45\textwidth]{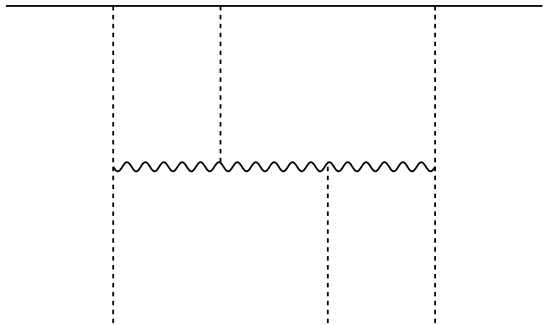}\\}
  \caption{\it Typical Regge-Gribov rescattering diagram, contributing at higher orders to the eikonal; dashed (wavy) lines denote exchanged (emitted) gravitons which have mostly transverse (longitudinal) propagators. 
    \label{5}}
\end{figure}

In the effective action framework, the elastic S-matrix of the tree diagrams in fig.~(\ref{2}) is given in terms of the classical solutions of the lagrangian equations of motion as
\eqn\label{action}
S(b, s)~&=&~\exp[\frac{i}{\hbar}\A(h^{\mu\nu}_{cl})];\\ \nn
\A(\tilde{h}^{++}, \tilde{h}^{--}, \Phi)~&=&~\int d^4x(\Lc_0+\Lc_e+\Lc_r+T_{++}\tilde{h}^{++} + T_{--}\tilde{h}^{--})\, ,
\eqnx
where $\Phi(x^+,x^-,\xt)$ generalizes $\phi(\xt)$ to four dimensions, and
\eq 
T_{--}=\kappa E\delta(x^-)\delta(\xt),~~T_{++}=\kappa E\delta(x^+)\delta(\xt-\bt)
\eqx
represent (up to an unconventional but convenient factor of $2\kappa$), the energy-momentum tensor of the colliding particles. The lagrangian consists of a kinetic term
\eq
\Lc_0=-\partial^*\tilde{h}^{++}\partial \tilde{h}^{--}+4\partial_+{\partial^*}^2\Phi\partial_-\partial^2\Phi^*\, ,
\eqx
where the longitudinal fields have a mostly transverse propagator and the (complex) $\Phi$ field a mostly longitudinal one, of a graviton emission term 
\eq
\Lc_e=\kappa(\mathcal{J}|\partial|^2\Phi^*+\mathcal{J}^*|\partial|^2\Phi);~~|\partial|^2\mathcal{J}=[{\partial^*}^2\tilde{h}^{++}\partial^2\tilde{h}^{--}-|\partial|^2\tilde{h}^{++}|\partial|^2 \tilde{h}^{--}]
\eqx
related to the reduced one in eq.~(\ref{reduced_a}), and, finally, of a rescattering term
\eq\label{rescatt}
\Lc_r=\kappa(\tilde{h}^{++}{\partial^*}^2\Phi^*{\partial_+}^2\partial^2\Phi+\tilde{h}^{--}\partial^2\Phi^*{\partial_-}^2{\partial^*}^2\Phi)\, ,
\eqx
which is supposed to take into account the rescattering diagrams of fig.~(\ref{5}). This term is quadratic in $\Phi$, and is likely to play a role when the latter is large. 

Here we do not treat the action (\ref{action}) in detail, but we would like to discuss a couple of important points. First of all, it was shown in \cite{ACV3} that the reduced action and equations of sec.~(\ref{s:model}) correspond to a shock-wave solution of the present lagrangian equations without rescattering terms, of the form
 \eqn\label{metric_LL}
\tilde{h}^{++}~&=&~\kappa \sqrt{s}~\delta(x^-)a(\xt),~~~\tilde{h}^{--}~=~\kappa\sqrt{s}~\delta(x^+)\bar{a}(\xt);\\ \nn
\Phi~&=&~\frac{\kappa^3s}{4}~\Theta(x^+x^-)\phi(\xt)\, ,
\eqnx
where now the longitudinal fields $a,\bar{a}$ and the transverse field $\phi$
appear as profile functions in front of the $x^+,x^-$ dependence. Note that, while the longitudinal part is of Aichelburg-Sexl type, the transverse part has support  inside the whole light-cone. This propagation, of retarded plus advanced type, corresponds to the principal value part of the Feynman propagator and is appropriate for the real part of the amplitude \footnote {Restoring the transverse propagation corrections to the $\Phi$ field solution amounts to effectively cutoff the wave inside the light-cone around $x^+x^-\lesssim R^2$ without modifying the wave-front, and thus the derivation of the reduced action.}.

 The shock-wave interpretation of the reduced action framework allows to embed it in spacetime, and to provide, in particular, the effective metric produced by the solutions of the lagrangian equations. The explicit $x^{\pm}$-dependence in eq.~(\ref{metric_LL}) allows to calculate the longitudinal components of the metric induced by the $\Phi$ field by generalizing eqs.~(\ref{FTA}) and (\ref{metric_TT})
to $x^{\pm}$-space and by using the longitudinal components of the $TT$ polarization
\eq
\epsilon_{++}^{TT}=-\frac{\partial_+}{4\partial_-},~~\epsilon_{--}^{TT}=-\frac{\partial_-}{4\partial_+},~~\epsilon_{+-}^{TT}=\frac{1}{4}\, .
\eqx
Taking into account the different normalization of  the metric components mentioned before, 
this procedure leads to the following expression for the induced metric in terms of $a$, $\bar{a}$ and a real-valued $\phi$:
\eqn\label{metric_tot}
ds^2&=& -dx^+dx^- (1- \frac12(\pi R)^2 \Theta(x^+ x^-) \nabla^2 \phi) + 2\pi R \left( a(z) \delta(x^-)(dx^-)^2+\bar{a}(z)  \delta(x^+)(dx^+)^2\right) \nn \\
&-& \frac14(\pi R)^2 \nabla^2 \phi \left( |x^+|  \delta(x^-)(dx^-)^2+
  |x^-|  \delta(x^+)(dx^+)^2 \right) + ds_T^2 \nn \\
  ds_T^2&=&|dz|^2+(\pi R)^2\Theta(x^+ x^-) (2|\partial|^2\phi~|dz|^2-\partial^2\phi~ dz^2-{\partial^*}^2\phi~{dz^*}^2)\\ \nn
&=& |dz|^2+(\pi R)^2\Theta(x^+ x^-) \left( \delta_{ij} \nabla^2- \nabla_i \nabla_j \right)\phi~ dx^i dx^j\, .
\eqnx
It is easy to check that the perturbation of the metric proportional to $\phi$ is transverse and traceless: it has  exactly the form of the $TT$ polarized gravitational field introduced in sec. (2.1), and is meant to describe the intermediate $h$ field contributing to the real part of the amplitude. A discussion of the features of the above  effective metric is postponed to sec.(\ref{s:$b>0$}), when explicit solutions will be available.

Note finally that we shall not consider in the following rescattering and string contributions to higher orders of the eikonal expansion. This is an acceptable approximation for the small $\phi$ regime $R\lesssim b$, but is likely to be insufficient when looking at distances smaller than $R$ and approaching the string length, when $\phi$ becomes large (cf. sec.~(\ref{s:axi})). Note that in this region rescattering and string effects are probably intertwined, because of the eikonal couplings occurring in the rescattering vertex of eq.~(\ref{rescatt}). They produce factors of $k^+k^-$ in the numerator of the corresponding diagrams and thus make the latter formally divergent, by emphasizing the role of large intermediate masses. Therefore, string excitations are required in order to regularize the sum over intermediate masses, and become non negligible. We have here a situation similar to normal diffractive excitation of string massive states by initial particles, where string corrections are taken into account, eventually, by the simple shift in eq.~(\ref{eikonal2}). In the present case it is the intermediate graviton to be excited in a sort of double-diffractive string excitation. We similarly hope that the present effects will turn out to be calculable, perhaps by introducing, in the framework of sec.~(\ref{s:model}), the Regge-graviton string emission vertices of ref~\cite{ABC}.

\section{Axisymmetric solutions}\label{s:axi}
\setcounter{equation}{0}

\subsection{Particle-particle  scattering  at $b=0$  }\label{s:paxi}

We start considering the reduced action of sec.~(\ref{s:model}) in the complementary region to the perturbative one we started with in sec.~(\ref{s:Hdiag}), by taking $\ls\ll b\ll R$. Since $b>\ls$ we shall not consider string corrections explicitly, even if the string has played an important role in assessing the validity of the model. This means -- since $R$ is the only explicit coupling being considered-- that we actually take the $b=0$ limit of a head-on collision. 
This is a fully non-perturbative regime whose interpretation requires  a non-trivial matching with the perturbative regime that we shall discuss in the following sections.

In the $b=0$ limit, we can look for axisymmetric solutions $a=\bar{a}$ and $\phi$ which are functions of $r^2=\xt^2$ only. Surprisingly, the nonlinear equations ~(\ref{reduced_eom}) take a particularly simple form that will allow a complete treatment of their solutions. Indeed, by setting $\dot{a}\equiv \partial a/\partial r^2$ etc.., they read, for $r \ne 0$,
\eqn\label{eom_a}
\frac{\partial}{\partial r^2}[r^2\dot{a}(1-(2\pi R)^2\dot{\phi})]~&=&~0,\\ \label{eom_b}
\frac{\partial}{\partial r^2}[r^2\ddot{(r^2\dot{\phi})}]+\frac{1}{2}\frac{\partial}{\partial r^2}(r^2\dot{a}^2)~&=&~0\, ,
\eqnx
thus providing, by inspection, two constants of motion with respect to the ``time'' variable $r^2$. The first one in eq.~(\ref{eom_a}) is fixed by the Gauss theorem on the delta function in~(\ref{reduced_eom}) to be $-1/2\pi$ and the other vanishes by the same token, so that we have  
\eqn\label{eom_1}
 r^2 \dot{a} (1-(2\pi R)^2\dot{\phi})=C_1=-\frac{1}{2\pi},~~~r^2(\ddot{r^2\dot{\phi}}+\frac{1}{2}\dot{a^2})~=~C_2~=~0 \, .
\eqnx

It is now convenient to introduce the function:
\eqn\label{rho}
\rho(r^2)\equiv r^2(1-(2\pi R)^2\dot{\phi}) \, ,
\eqnx
  which has dimension of a squared length, and embodies the effect of the (transverse) emission field which will play an important role in the following. By expressing $\dot{\phi}$ and $\dot{a}$ in eq.~(\ref{eom_1}) in terms of $\rho$ we have
\eqn\label{eom_2_a}
\dot{a}(r^2)~&=&~-\frac{1}{2\pi \rho(r^2)},\\ \label{eom_2_b}
\ddot{\rho}(r^2)-\frac{R^2}{2\rho^2}~&=&~0,~~~\dot{\rho}^2+\frac{R^2}{\rho}~=~C_3\, ,
\eqnx
where the constant $C_3$ will be determined by requiring consistency with the perturbative expansion for $r^2\gg R^2$. Indeed, in this limit, the system of eqs.~(\ref{reduced_eom}) reduces to eqs.~(\ref{a_0}) and (\ref{h_0}) thus showing, by (\ref{Hbypol}), the large-$r$ behaviour
\eqn
a(r^2)=\bar{a}(r^2)~\simeq ~-\frac{1}{2\pi}~\log r^2;~~r^2\dot{\phi}~\simeq ~\frac{1}{8\pi ^2}\log r^2,~~\rho(r^2)\simeq r^2-\frac{R^2}{2}\log~r^2 \, ,
\eqnx
which implies $\dot{\rho}\rightarrow 1$ at large distances and thus $C_3=1$ in eq.~(\ref{eom_2_b}), yielding finally
\eq\label{eom_3}
\dot{\rho}^2+\frac{R^2}{\rho}~=~1,~~~\dot{a}~=~-\frac{1}{2\pi \rho}\, .
\eqx

We conclude that our nonlinear problem reduces to the classical Coulomb problem for the ``radius'' $\rho(r^2)$ at ``time'' $r^2$. Since the ``Coulomb potential'' in eq.~(\ref{eom_3}) is repulsive, we can say from start that, coming from a positive $\rho(r^2)\sim r^2$ at large distances, the generalized radius $\rho(r^2)\ge R^2$ will never vanish during $r^2$-evolution, even at $r^2=0$. This means that $\dot{a}$ in eq~(\ref{eom_2_a}) is  not singular at $r^2=0$ -- contrary to its perturbative behaviour -- and that, by eq.~(\ref{eom_1}), $(2\pi R)^2\dot{\phi}\simeq -\rho(0)/r^2$ must be {\it singular} instead, a feature which is non-perturbative as well. 

The above non-perturbative behaviour of $\dot{a}$ and $\dot{\phi}$ is somewhat puzzling. It means that in the first eq.~(\ref{eom_1}) the outgoing flux of $\nabla a$ is traded for that of $\nabla\phi$, with increase of the $\phi$ field at small distances and a non-vanishing value of $r^2\dot{\phi}$, due to $\rho(0)\ne 0$. It is caused by the large curvature $\ddot\rho$ for small values of $\rho$ which forbids $\rho(0)=0$ for non-negative, real-valued solutions. On the other hand, the condition $\rho(0)=0$ is not only a property of the perturbative behaviour, but appears to be {\it required} in order to avoid or to treat properly a boundary term at $r=0$ in the reduced action, as discussed in the Appendix. 
We shall see that only for $b\gtrsim R$ we will be able, in the next section, to meet that condition for real-valued solutions, thus obtaining a nonsingular $\dot{\phi}$. Alternatively, we can give up the reality requirement suggested by the metric interpretation, and look for complex  $b=0$ solutions. We shall motivate and explore such possibility in sec. 7.

The explicit solutions of eqs.~(\ref{eom_3}) for $\rho$ and $\bar{a}=a$ are obtained by standard methods in terms of a hyperbolic angle $\chi(r^2)$ as follows
\eqn\label{soln_a}
\rho(r^2)~&=&~R^2\cosh^2\chi(r^2);~~r^2=R^2(\chi+\cosh\chi~\sinh\chi-\chi_0-\cosh\chi_0~\sinh\chi_0)\\ \label{soln_b}
a(r^2)~&=&~\frac{1}{2\pi}\int_{r^2}^{L^2}\frac{dr^2}{\rho(r^2)}=\frac{1}{\pi}(\chi(L^2)-\chi(r^2))\, ,
\eqnx
where $\chi_0=\chi(0)$ is the arbitrary value of $\chi$ at the origin.
Its presence is not surprising, because we have set only one boundary condition, providing matching to perturbation theory at large distances ($\dot{\rho}(\infty)=1$). The explicit solution (\ref{soln_a}) shows that the boundary condition  $\rho(0)=0$ can only be met by a complex value of $\chi_0$ (like $\chi_0=-i\pi /2$) -- a case that will be discussed in detail in sec. 7. For the moment we treat $\chi_0$ as a free parameter, even if a way to determine a real value for it will be discussed in the next section, as an alternative to $\rho(0)=0$ for $b\lesssim R$. 

Since $\chi(r^2)$ is a monotonically increasing function, there are two kinds of real-valued solutions, depending on the sign of $\chi_0$. 
If $\chi_0>0$, $\rho(r^2)$ increases monotonically to $\rho\sim r^2$ at large distances, while for $\chi_0<0$ $\rho$ decreases first to its minimum $\rho=R^2$ -- corresponding to $\chi=0$ -- and then increases to $\infty$. The scale of the large-$r^2$ behaviour of $\rho$ is itself dependent on $\chi_0$. Indeed, a simple iterative evaluation of eq.~(\ref{soln_a}) yields the more detailed behaviour
\eqn\label{asy_a}
\rho(r^2)~&\simeq&~r^2-\frac{R^2}{2}\log\frac{4r^2}{\bar{r}^2(\chi_0)};~~
\bar{r}^2(\chi_0)~=~R^2\exp(1+2\chi_0 + \sinh 2\chi_0)\\ \label{asy_b}\dot{\phi}~&\simeq&~\frac{1}{8 \pi^2 r^2}\log\frac{4r^2}{\bar{r}^2(\chi_0)};~~\phi~\simeq~\frac{1}{16\pi^2}\log^2{\frac{4r^2}{\bar{r}^2(\chi_0)}} \, ; \, \\ a(r^2)~&=&~ \frac{1}{2\pi} \left( \log \frac{L^2}{r^2} + \frac{R^2}{2 r^2} \log \frac{4r^2}{\bar{r}^2} \right)\, ,
\eqnx
which is actually valid for any value of $\chi_0$.

Let us remark that the arbitrary constant occurring in the integration of (\ref{eom_3}) for $a(r^2)$ has been traded, in eq.~(\ref{soln_b}), for a scale $L$, defined by $a(L^2)=0$, that will play the role of infrared scale, as in eq.~(\ref{a_0}). The latter is then fixed to be the same IR cutoff $L\gg R$ needed for the evaluation of the action below, so that $a(0)$ carries the large logarithm $\chi(L^2)\simeq \log \frac{2L}{R}$. 

The reduced action of eq.~(\ref{action_os}) can be explicitly evaluated  on the axisymmetric solutions, and takes the form
\eqn\label{action0}
\frac{\A(0,s)}{2\pi Gs}~&=&~a(0)+\bar{a}(0)- 2\pi\int_{0}^{L^2}dr^2~r^2~\dot{a}^2~+~2\pi (2\pi R)^2\int dr^2~[\frac{d(r^2\dot{\phi})}{dr^2}]^2\\ \nn
&=&2a(0)+I_a+I_{\phi}\, ,
\eqnx 
where the integral $I_a$ is IR divergent, but can be made finite by combining it with $a(0)$ as follows
\eqn\label{IaIphi0}
I_a+a(0)&=&
\frac{1}{2\pi}\int_{0}^{L^2} \frac{dr^2}{\rho}(1-\frac{r^2}{\rho})=\frac{1}{2\pi}(1-e^{-2\chi_0});\\ \nn
I_{\phi}&=&\frac{1}{2\pi R^2}\int_0^{\infty}dr^2(1-\dot{\rho})^2=\frac{\exp(-2\chi_0)}{2\pi}\, ,
\eqnx
where the finite $L\rightarrow \infty$ limit of $I_a+a(0)$ and $I_{\phi}$ have been easily evaluated by use of eqs.~(\ref{soln_a}) and (\ref{soln_b}). After simple algebra we obtain
\eq\label{action0}
\A(0,s)=Gs~(2\pi a(0) +1)=2Gs~(\chi(L^2)-\chi_0 +\frac{1}{2})\simeq 2Gs~(\log{\frac{2L}{R}}-\chi_0 +\frac{1}{2})\, ,
\eqx
which shows a simple additive dependence on $\chi_0$ and the expected IR divergent Coulomb phase. The latter is unrenormalized, and can be factorized away in the S-matrix, as usual.

Inserting the solution (\ref{soln_a}, \ref{soln_b})  in Eq. (\ref{metric_tot}) we obtain an explicit  expression for the metric  where we note the appearance, besides that of $\Delta \phi$, of the field $\dot \phi$ that -- if $\rho(0)\ne 0$ -- generates the behaviour $r^{-2}$ in some metric components. 
Further comments on this issue  and a discussion of the effective metric will be given after having extended our analysis to generic values of $b$.

\subsection{Central collision of two homogeneous beams }\label{s:baxi}

    An interesting case that can also be solved analytically but, unlike the previous one,  contains a tunable parameter is that of the central  collision of two homogeneous, finite-size beams of massless particles.
The point here is  that our effective-action method should retain its valididity even when the point-like sources are replaced by smooth (null) energy distributions on the transverse plane.
A particularly simple case is that of two circular homogeneous  beams of radius  $\Sigma$ (area $\pi \Sigma^2$), each one carrying a total amount $E$ of energy, and undergoing a head-on collision. The problem is again axisymmetric and is characterized by the dimensionless parameter:
\eqn
\frac{R}{\Sigma} = 4 \frac{G E}{\Sigma} =  4 \pi G \epsilon \Sigma = \frac{\Sigma}{2 f}\, ,
\eqnx 
where $\epsilon$ is the energy density per unit area and $f = (8 \pi G \epsilon)^{-1}$ is the
focal distance for geodesics impinging on the beam-shaped shock wave (see e.g. \cite{FPV}).

At {\it classical} GR level, the problem of determining  when a closed trapped surface (CTS) is produced by the collision led to  the conclusion  \cite{KV} that  a CTS forms when the  above ratio exceeds a critical value, for which an upper limit was established:
\eqn
\label{CTSB}
\left( \frac{R}{\Sigma}\right)_{CTS}  =  \left(\frac{\Sigma}{2 f}\right)_{CTS} < 1\, .
\eqnx 

It is quite  interesting to investigate this problem within  our  present quantum approach. It is straightforward to adjust our effective action equations (\ref{eom_1})  to this new situation. They get simply modified as follows:
\eqn\label{eom_1beam}
 r^2 \dot{a} (1-(2\pi R)^2\dot{\phi}) =  - \frac{1}{2\pi}\theta(r-\Sigma) - \frac{r^2}{2\pi\Sigma^2} \theta(\Sigma-r) 
  ,~~~\ddot{r^2\dot{\phi}}+\frac{1}{2}\dot{a^2}=0\, ,
 \eqnx
from which, using again $\rho$ as defined in (\ref{rho}), we obtain
\eqn\label{eom_2_beam}
\ddot{\rho}(r^2)= \frac{R^2}{2\rho^2} \theta(r-\Sigma) +
 \frac{R^2 r^4}{2 \Sigma^4 \rho^2} \theta (\Sigma-r) \, .
  \eqnx

In other words, the equation for $\rho$ is unchanged at large $r >\Sigma$ but is strongly modified (though in a continuous way) for  $r < \Sigma$.
At small $r$ the equation has a regular solution  with $\rho(0)=0$ and a nice, analytic  expansion around  $r^2=0$, which can be computed after inserting some value for $\dot{\rho}(0)$. This solution, however, should match the one from $r > \Sigma$ at the $r= \Sigma$ boundary. In this  latter solution one has, as before,
\eqn\label{eom_3_beam}
\dot{\rho} = +\sqrt{ 1-  \frac{R^2}{\rho}} \Rightarrow  \dot{\rho}(\Sigma^2) = +\sqrt{ 1-  \frac{R^2}{\rho(\Sigma^2)}}=\tanh\chi_{\Sigma},~~~(\chi_{\Sigma}\equiv\chi(\Sigma^2))   \, ,
  \eqnx
and thus the initial condition on $\dot{\rho}(0)$ has to be chosen so as to satisfy (\ref{eom_3_beam}).
This, however, turns out to be impossible  if:
\eqn\label{eom_4_beam}
\Sigma^2 \dot{\rho}(\Sigma^2) =  \Sigma^2 \tanh\chi_{\Sigma}   < \rho(\Sigma^2) = R^2 \cosh^2\chi_{\Sigma} \, ,
  \eqnx
simply because the concavity of the $\rho$-curve (due to  $\ddot{\rho}(r^2) \ge 0$) will prevent such a curve to pass through the origin, similarly to the previously discussed  $b=0$ case.
Such a simple concavity argument gives an upper limit on the value of $R/\Sigma$ for which the condition $\rho(0)=0$ can be imposed for real-valued solutions.  Equivalently, it gives a lower bound for the critical value of $\Sigma$, $\Sigma_c$, for that to happen. It  is easily computed to be given by $(R/\Sigma)_c<2^{1/2}3^{-3/4}\simeq 0.62$.
              
On the other hand, we can also provide an upper bound on $\Sigma_c$, and thus prove the existence of two distinct regimes, by noting that, by eq.~(\ref{eom_2_beam}), 
\eq\label{crit_beam}
0 < \Sigma^2\dot{\rho}(\Sigma^2)-\rho(\Sigma^2)~=~\frac{R^2}{2\Sigma^4}\int_0^{\Sigma^2}dr^2\frac{r^6}{\rho^2(r^2)}<\frac{R^2}{4\dot{\rho}^2(0)}\simeq \frac{R^2}{4\dot{\rho}^2(\Sigma^2)}\, .
\eqx
By replacing in (\ref{crit_beam}) the values of the solution for $r>\Sigma$, we get the relation
\eq
0 < \Sigma^2 \tanh\chi_{\Sigma}- R^2 \cosh^2\chi_{\Sigma}\lesssim \frac{R^2}{4\tanh^2\chi_{\Sigma}} \, ,
\eqx
where we realize that the r.h.s. is of order $R^2$, without any particular enhancement when $\Sigma\gtrsim R$ increases, so that a solution for $R/\Sigma_c$ can be found. Therefore, for $\Sigma>\Sigma_c$, the condition $\rho(0)=0$ can be met by real-valued field solutions.

In order to get a more precise estimate of $\left( R/\Sigma\right) _c $ we have solved numerically the differential equation (\ref{eom_2_beam}) and looked for a critical value above which it is no longer possible to impose the condition $\rho(0)=0$. The result of such an analysis gives:
 \eqn
 \left( R/\Sigma \right) _c \sim 0.47 \, ,
 \eqnx
  in line with the classical CTS-bound (\ref{CTSB}).

 The above discussion suggests a (loose?) correspondence between ``untrapped'' classical GR solutions and our real-valued field solutions  that satisfy $\rho(0)=0$ and match  perturbation theory at large distances. In either case such solutions cease to be available when the beam size is smaller than some critical radius below which gravitational trapping occurs on the classical side, and the small-$r$ $\dot{\phi}$- singularity develops on the other. We shall find a similar phenomenon in the case of particle-particle scattering at $b>0$, to be discussed next.

\section{Extension to $b>0$ and critical impact parameter}\label{s:$b>0$}
\setcounter{equation}{0}

For nonvanishing $b$ the solutions, of course,  are not axisymmetric and show a nontrivial azimuthal dependence on $\theta=\theta_{\bt\xt}$. We shall simplify the issue by performing an azimuthal average on $\theta$ and, furthermore, by performing a spin-$0$ projection of the $\bar{a}\leftrightarrow a$ relationship. Instead of the simple translation $\xt \rightarrow (\bt-\xt)$ we shall take the relation (better expressed for the Fourier transform $\tilde{a}(\kt)$)
\eq\label{aabar}
\bar{a}(b,\xt)=\int d[\kt]\tilde{a}(\kt)\exp(i\kt\xt)J_0(b|\kt|)\, ,
\eqx
which has no memory of the direction of $\bt$ and is equivalent to an azimuthal average if $\tilde{a}$ only depends on $\kt^2,\bt^2$. Note that this procedure singles out the vector $\xt$ corresponding to the source at $\xt=0$ with respect to the vector $\bt-\xt$ and is therefore asymmetrical with respect to $a$ and $\bar{a}$. We can interpret it by saying that we look at the $\xt^2$-dependence of  $a$ in the average field of $\bar{a}$, which is much similar to the collision of a point-particle with a ring-shaped source. Then, the symmetrical result will be obtained if we look at the $(\bt-\xt)^2$-dependence of $\bar{a}$ in the average field of $a$. 

By replacing eq.~(\ref{aabar}) in eqs.~(\ref{reduced_eom}) we soon realize that the ansatz $a=a(\xt^2 = r^2)$, $\bar{a}= \bar{a}(r^2)=\la a((\bt-\xt)^2)\ra_{\theta}$ is self-consistent and, by the same manipulations as in sec.~\ref{s:axi}, we obtain the equations
\eqn  \label{axib}
\Dr[r^2 (\Dr)^2(r^2\dot{\phi})]~&=&~-\frac{1}{2}\Dr(r^2\dot{a}\dot{\bar{a}}),\\ \nn
\Dr[r^2\dot{a}(1-(2\pi R)^2\dot{\phi})]~&=&~0~~~
\eqnx
and, therefore, using again the function $\rho$ of (\ref{rho}):
\eq \label{axibconst}
\dot{a}\rho(r^2) =-\frac{1}{2\pi},~~~\ddot{\rho}= 2 (\pi R)^2 \dot{a}\dot{\bar{a}} \, .
\eqx

\subsection{Solutions for $b>0$ and perturbative expansion}

Eqs.~(\ref{axibconst}) differ from those valid at $b=0$ by the replacement of a factor of $\dot{a}$ by $\dot{\bar{a}}$. In order to relate $\bar{a}=\la a((\bt-\xt)^2)\ra_{\theta}$ to $a$ let us note that eq.~(\ref{aabar}) has two distinct regimes, according to whether $r\gg b$ ($r\ll b$). In such regimes one can set, approximately, $b=0$ ($r=0$) in the right hand side. Therefore, we are led to replace $\bar{a}$ with the simple approximation
\eq\label{aabar_appro}
\bar{a}(r^2)~\simeq~ a(r^2)\Theta(r^2-b^2)+a(b^2)\Theta(b^2-r^2);~~\dot{\bar{a}}(r^2)~\simeq~\Theta(r^2-b^2)\dot{a}(r^2)  \, .
\eqx
  Note that this approximation is exact for $a_0$ of eq.~(\ref{a_0}) and for the collision with the ring-shaped source envisaged before.
    
Introducing  the above approximation in eqs.~(\ref{axibconst}) one has
\eq
\ddot{\rho}(r^2)~=~\frac{R^2}{2\rho^2} ~\Theta(r^2-b^2)\, ,
\eqx 
so that, for $r^2<b^2$, the ``repulsive'' Coulomb potential is absent and $\ddot{\rho}=0$. This in turn leads to the solution
\eqn\label{axisolnb}
\rho~&=&~R^2\cosh^2\chi(r^2),~~~(r^2 > b^2);~~\rho~=~\rho(b^2)+\dot{\rho}(b^2)(r^2-b^2),~~(r^2\le b^2);\\ \nn
r^2~&=&~b^2+R^2(\chi+\sinh\chi~\cosh\chi-\chi_b-\sinh\chi_b~\cosh\chi_b)\, ,
\eqnx
where we have introduced the hyperbolic angle $\chi=\chi(r^2,b)$ and the notation $\chi_b\equiv\chi(b^2)$.
The corresponding longitudinal and transverse fields are
\eqn\label{axifieldsb}
a(r^2)~&=&~\frac{1}{2\pi}\int_{r^2}^{L^2}\frac{dr^2}{\rho(r^2)}=\frac{1}{\pi}(\chi(L^2)-\chi(r^2))~~~(r\ge b)\\ \nn
&=&~\frac{1}{\pi}(\chi(L^2)-\chi_b)+\frac{1}{2\pi t_b}\log\frac{\rho(b^2)}{\rho(0)+t_br^2}~~~(r<b)\\ 
h_{TT}=h(r^2)&=&4|\partial|^2 \phi=\frac{1-\dot{\rho}}{(\pi R)^2}=\frac{1-\tanh \chi(r^2)}{(\pi R)^2}\, ,
\eqnx
where $t_b\equiv\tanh\chi_b$, $\rho(0)=\rho(b^2)-b^2t_b$, and we have fixed the additive constant in the longitudinal field by requiring $a(L^2)=0$, $L$ being the IR cutoff parameter.

Because of the linear behaviour of $\rho(r^2)$ for $r^2<b^2$, eq.~(\ref{axisolnb})
leads to the possibility of enforcing the boundary condition $\rho(0)=0$, typical of the perturbative expansion, and required for self-consistency by the reduced action itself (see Appendix). For that to happen we must have
\eq\label{critical}
\rho(b^2)=R^2 \cosh^2\chi_b=b^2\dot{\rho}(b^2)=b^2\tanh\chi_b;~~~\frac{R^2}{b^2}=t_b(1-t_b^2)\, ,
\eqx
a condition which resembles eq. (\ref{crit_beam}) found before.

The {\it criticality equation} ~(\ref{critical}) is cubic in the $t_b$ parameter and determines the branches of possible solutions with $\rho(0)=0$. At the critical value $b_c^2=3\sqrt{3}R^2/2$ of the impact parameter, the equation is stationary. For $b>b_c$, there are two solutions with nonnegative $\rho$, one with $t_b\rightarrow 1$ for $b\gg b_c$ - which will be related to the perturbative one - and the other with $t_b \rightarrow 0$. The third formal solution,with $t_b<-1$ is actually to be discarded because it would require $\rho(r^2)<0$ at large distances as well. For $b \gg b_c$, the solution with larger $\chi_b\rightarrow \infty$ matches the perturbative solution at large distances. In fact, a simple iterative evaluation of eq.~(\ref{axisolnb}) yields the large-$r$ behaviour 
\eq
\rho(r^2)~\simeq~r^2-\frac{R^2}{2}\log\frac{4r^2}{\bar{r}^2(\chi_b)}:~~
\bar{r}^2(\chi_b)=R^2\exp(1+2\chi_b +\sinh2\chi_b-\frac{2b^2}{R^2})\, .
\eqx
Since $\exp2\chi_b \simeq 4b^2/R^2$ for $b\gg b_c$, it follows that $\bar{r}^2(\chi_b)/4 \simeq b^2 $ is just the scale of the perturbative solution in sec.~\ref{s:Hdiag}, as anticipated.

On the other hand, for $b<b_c$, there are no real valued solutions to eq.~(\ref{critical}) with $\rho(b^2)\ge 0$ nor, equivalently, to the boundary condition $\rho(0)=0$. It is not clear how to replace this boundary condition and thus to define a meaningful real valued solution for $b<b_c$. For instance -- since we cannot reach $\rho(0)=0$ -- we can try to do our best and look for a $\chi_m(b)$ such that $\rho(b^2)-b^2\dot{\rho}(b^2)$ is minimal. This yields the condition
\eq
\frac{b^2}{2R^2}=\cosh^3\chi_m~\sinh\chi_m=\frac{t_m}{(1-t_m^2)^2}\, ,
\eqx 
which, for any $b<b_c$, admits real solutions such that, while $b$ decreases, $\chi_m$ decreases from $\chi_m(b_c)=\chi_c$ to $\chi_m(0)=0$ and $\rho(0)$ increases from $0$ to $R^2$. Therefore, this kind of solution determines $\chi_0=0$ as its $b=0$ limit. However, there appears to be no compelling reason for this choice, except perhaps that, among the real-valued solutions,  the ``distance'' of this one to the $\rho(0)=0$ complex solution to be studied in sec.7 is smallest.

We remark that in this $b>0$ case, like in the problem discussed in sec. 3.2,   the critical impact parameter $b_c$ separates -- in the real-valued domain -- the class of ``weak-field'' solutions having $\rho(0)=0$ (for $b>b_c$) from that of ``strong-field'' solutions with $\rho(0)>0$ and a small-$r$ $\dot{\phi}$-singularity (for $b<b_c$). The latter solutions, however, appear to be somewhat ill-defined.

\subsection{The on-shell action and its singularities}
By evaluating the action in eq.~(\ref{reduced_a}) on the solutions (\ref{axisolnb}) we find the expression
\eqn
\frac{\A(b,s)}{2\pi Gs}~&=&~a(b^2)+\bar{a}(0)-\frac{1}{2}\int d^2\xt \nabla \bar{a}\nabla a +\frac{(\pi R)^2}{2}\int d^2\xt(\nabla^2\phi)^2 \\ \nn
&=&2a(b^2)+I_a(b)+I_{\phi}(b)\, ,
\eqnx
where we have used $\bar{a}(0)=a(b^2)$ and we have evaluated the integrals 
\eqn\label{IaIphi}
I_a(b)+a(b^2)&=&\frac{1}{2\pi}\int_{b^2}^{\infty} \frac{dr^2}{\rho(r^2)}\left(1- \frac{r^2}{\rho(r^2)}\right)=\frac{1}{2\pi }\left(1-e^{-2\chi_b}-\frac{2b^2e^{-\chi_b}}{R^2\cosh{\chi_b}}\right); \nn \\ 
I_{\phi}(b)&=&\frac{1}{2\pi R^2}\int_0^{\infty}dr^2(1-\dot{\rho})^2=\frac{\exp(-2\chi_b)}{2\pi}\left(1+\frac{b^2}{R^2\cosh^2{\chi_b}}\right)\, .
\eqnx
In the actual evaluation, we find that the integrals (\ref{IaIphi}) are related by an integration by parts yielding $I_{\phi}=-(I_a+a(b^2))/2$, except for a  boundary contribution $\sim\rho(0)(1-\dot{\rho}(0))$. The latter is discussed in more detail in the Appendix, where we argue that its consistent treatment would require $\rho(0)=0$ in all instances, even at the cost of picking up complex solutions of the field equations. Nevertheless, if we decide to keep it, we obtain, after simple algebra
\eq\label{actionb1}
\A(b,s)~=~Gs~\left( 2(\chi(L^2)-\chi_b) +1-\frac{b^2}{R^2\cosh^2{\chi_b}}\right)~~~~\, .
\eqx

It is amusing to note that the above expression for the action is stationary with respect to  $\chi_b$ at fixed $b$ precisely when the ``criticality condition"   (\ref{critical}) holds. This suggests an alternative interpretation of the condition $\rho(0)=0$, namely that of requiring stationarity  in a ``sum over solutions'' (or perhaps better over collective coordinates contained in the solutions) definition of the S-matrix
\eq\label{chi_integral}
S(b, s)~=~\int d\chi_b \mu(\chi_b)\exp(i\A (b, s;\chi_b))\, ,
\eqx
where however the integration measure $\mu(\chi_b)$, possibly related to a functional fluctuation determinant around the given solution, is actually not available.

Sticking for the moment to real-valued solutions, and using, for $b>b_c$, the criticality equation in order to eliminate $b^2$ in terms of $\chi_b$, we finally get the convenient  expression
\eq\label{actionb2}
\A(b,s)~=~Gs~\left(2(\chi(L^2)-\chi_b)+1-1/t_b\right);~~~(b>b_c)\, .
\eqx
By evaluating $\chi_b$ from (\ref{critical}) we find the large-$b$ behaviour $\A \simeq 2Gs(\log\frac{L}{b} +\frac{R^2}{4b^2})$, which checks with the perturbative expansion, the slight difference of the $R^2/b^2$-correction being due to the azimuthal averaging procedure.

The known $b$-dependence of the action allows to find the elastic scattering amplitude $A(s, \qt^2)$ by a Fourier transform, and the related classical deflection angle by a stationarity equation in $\bt$:
\eqn\label{amplitude}
\frac{1}{s}A(s,\qt^2)~&=&~-2i\int d^2\bt \exp i\A(b, s)~\exp i\bt\qt \\ \label{angle1}
\qt_{cl}~&=&~-\nabla_{\bt} \A(b, s) \, .
\eqnx
By applying eq.~(\ref{angle1}), a simple calculation on eq.~(\ref{actionb1}) yields the expression of the deflection angle
\eq\label{angle2}
 \sin\frac{\theta_{cl}}{2}=\frac{q_{cl}}{\sqrt s}~=~\frac{R b}{\rho(b^2)}=\frac{R}{b\tanh \chi_b}\simeq \frac{R}{b}(1+\frac{R^2}{2b^2}+...)~~~(b\ge b_c)\, ,
\eqx
which determines the (resummed) corrections to the Einstein deflection as function of $b$, starting from large impact parameters down to the critical radius $b=b_c$. Actually, the deflection becomes maximal ($\theta_{cl}=\pi$) at some $b>b_c$, the limiting value of the r.h.s. of (\ref{angle2}) being $(4/ 3)^{1/4}>1$, showing that at such large angles the impact parameter framework is not fully meaningful. Alternatively, if we interpret $\A(b,s)$ as $\A(J=b\sqrt{s}/2, s)$, and we transform back to energy and scattering angle by convoluting $\A$ with $P_J(cos \theta)$, we arrive at the (perhaps more physical) result:
  \eq
 \theta_{cl} (b= b_c) =  2~(4/3)^{1/4} > \pi/2\, ,
 \eqx
 meaning that, at $b=b_c$, the two particles already invert  the sign of their relative momentum. 
 
Note  that the action (\ref{actionb2}) develops a branch cut singularity at $b^2=b_c^2=6\sqrt{3}~G^2s$. It is soon realized, using (\ref{actionb2}),  that, while $t_b$ has a square-root singularity at $t_b=t_c$, the action branch-cut is of type $\sim (b-b_c)^{3/2}$, because the total derivative $d\A/d\chi_b$ vanishes also at $\chi_b=\chi_c$, being $\sinh^2 \chi_c=1/2$. In other words, the action is stationary in $\chi_b$ at that point, just like the criticality equation (\ref{critical}). We expect this feature to be even more general than the present model, because, were the action stationary close to $\A_c$ but at a different value, there would be another pinch of two solutions besides the one we know at $\chi_b=\chi_c$. For the two to coincide, the action expansion around $\A_c$ should start at order $(\chi_b-\chi_c)^2$, the first nonanalytic piece being $(\chi_b-\chi_c)^3$. The action for $b \rightarrow b_c^+$ turns out to have the following expansion 
 \eqn \label{expansion_a}
\frac{ \A -\A_c}{G s}~& =&~ - 2\sqrt{3} ( \chi_b - \chi_c)^2 + \frac{16}{3} ( \chi_b - \chi_c)^3 + O(( \chi_b - \chi_c)^4) \\ \nn
 ~&=&~ \sqrt{3}(1-\frac{b^2}{b_c^2}) + \frac{2\sqrt{2}}{3} (\frac{b^2}{b_c^2}-1)^{3/2} \, ,
 \eqnx
 where the analytic piece  dominates around $b\simeq b_c$ and provides the deflection we have just discussed.

\subsection{The effective metric}
The metric describing Re$\delta$ and the reduced action is obtained from eq.~(\ref{metric_tot}) by specializing to the axisymmetric solutions $a(r^2)$ and $\phi(r^2)$, which are exact at $b=0$ and averaged out at $b>0$. We then obtain
\eqn\label{metric_b}
ds^2&=&-dx^+dx^- (1- 2(\pi R)^2 \Theta(x^+ x^-)~\frac{\partial (r^2\dot{\phi})}{\partial r^2}) + 2\pi R \left( a(r^2) \delta(x^-)(dx^-)^2+\bar{a}(r^2)  \delta(x^+)(dx^+)^2\right)\nn  \\ 
&-& (\pi R)^2 \frac{\partial (r^2\dot{\phi})}{\partial r^2} \left( |x^+|  \delta(x^-)(dx^-)^2+
  |x^-|  \delta(x^+)(dx^+)^2 \right) + ds_T^2 \nn \\
ds_T^2 &=&(1+2(\pi R)^2\Theta(x^+ x^-)\dot{\phi})dr^2+r^2 (1+2(\pi R)^2\Theta(x^+ x^-)(\dot{\phi}+2r^2\ddot{\phi}))d\theta^2 \, ,
\eqnx
where, for $b>0$, the azimuthal averaging is done at fixed $\xt^2\equiv r^2$.

We think that the effective metric so defined is really meaningful on the real valued lagrangian solutions for $b>b_c$ only. In fact, in such a case, we have matched the solution with larger $\chi_b$ to the perturbative expansion at large distances and, furthermore, the fields of eqs.~(\ref{axisolnb}) and (\ref{axifieldsb}) are well behaved in the small-$r$ region also. In particular, in the transverse part of the metric, the fields
\eqn\label{phifields}
\dot{\phi}~\simeq~-(\dot{\phi}+2r^2\ddot{\phi})~\simeq\frac{1}{8 \pi^2 r^2}\log\frac{4r^2}{\bar{r}^2(\chi_b)}~~~(r\gg R)
\eqnx   
have the role of decreasing the circumference over radius ratio at large distances, while $\dot{\phi}=1-t_b$ becomes just a constant at small distances.

On the other hand, if we take the real-valued solution defined above for $b<b_c$, the situation does not change much at large distances but, at short distances, the fields of eq.~(\ref{phifields}) both develop a $-\rho(0)/r^2$ singularity whose interpretation is doubtful, because of the rescattering and string corrections neglected in the present approach. Should we take that singularity seriously, the coefficient of $dr^2$ would become negative at some $r\ll R$ without any major change in the rest of the metric, because $d(r^2\dot{\phi})/dr^2$ is instead regular, except possibly at $r=0$. This is perhaps one more reason to stick to the condition $\rho(0)=0$ for $b<b_c$ also, even if that means considering complex solutions (cf. sec.7).

\section{Momentum space formulation}
\setcounter{equation}{0}

 \subsection{Effective action and equations of motion  }
 
 In order to reformulate the problem in momentum space and to make some symmetries more manifest it is convenient to place the sources for the fields $a$ and ${\bar a}$ at some generic points
 in transverse space $\bt_1$ and $\bt_2$. Normalizing Fourier transforms as:
 \begin{equation}\label{FTN}
a(\kt) = \int d\xt e^{i\kt \xt} a(\xt) \; \Rightarrow \;   a(\xt) = \frac{1}{4 \pi^2} \int d^2\kt e^{-i\kt \xt} a(\kt) \equiv \int [d \kt] e^{-i\kt \xt} a(\kt)\, ,
\end{equation}
the reduced effective action (\ref{reduced_a}) for the IR-safe polarization and after neglecting rescattering can be rewritten in momentum space as follows:
\eqn
\frac{\pi A}{Gs} &=&  \int \frac{d^2\kt}{\kt^2} \left[ e^{i\kt (\bt_1-\bt_2)} \gamma_1(\kt) +  e^{i\kt (\bt_2-\bt_1)} \gamma_2(\kt)-  e^{i\kt (\bt_1-\bt_2)} \gamma_1(\kt) \gamma_2(-\kt)  \right]  \nn \\
 &-& \frac{(\pi R)^2}{2} \int d^2 \kt  \left[  \frac{1}{2} h(\kt)  h(-\kt) -  h(-\kt) \Hc(\kt)  \right]  \, .
\eqnx
Here  $\gamma_i(\kt)$ and $h(\kt)$ are related to the Fourier transforms of  $a$, ${\bar a}$ and $h(\xt)$ by:
\begin{equation}
\gamma_1(\kt) = \frac{k^2 a(\kt)}{2}e^{-i\kt \bt_1} ~~~;~~~ \gamma_2(\kt) = \frac{k^2 {\bar a}(\kt)}{2}e^{-i\kt \bt_2}~~,~~
h(\kt) = -  k^2  \phi(\kt) \, ,
\end{equation}
and
\begin{equation}
 \Hc(\kt) \equiv  \frac{1}{\pi^2 \kt^2} \int d^2\kt_1 d^2\kt_2 \delta(\kt-\kt_1-\kt_2) \gamma_1(\kt_1)  \gamma_2(\kt_2)  
  e^{i(\kt_1\bt_1+ \kt_2 \bt_2)} \sin^2 \theta_{12}\, .
 \end{equation}
The equations of motion that follow from this action read:
\eqn \label{h=H}
h(\kt) &=& \Hc (\kt) \rightarrow  h_{TT}(\kt)   
~~ {\rm for} ~\bt_1 \rightarrow 0 , ~ \bt_2 \rightarrow \bt \\
\gamma_1(\pt) &=&  1 + \frac{R^2}{2} p^2  \int \frac{d^2\kt}{\kt^2} h(-\kt) \gamma_1(\kt+ \pt) e^{i\kt\bt_1} \sin^2\theta_{p/k+p}   \\
\gamma_2(\pt) &=&  1 + \frac{R^2}{2} p^2  \int \frac{d^2\kt}{\kt^2} h(-\kt) \gamma_2(\kt+\pt) e^{i\kt\bt_2} \sin^2\theta_{p/k+p} \, .
 \eqnx
  Eliminating $h(\kt)$ through (\ref{h=H}) we get  two coupled equations involving just
  $\gamma_1$ and $\gamma_2$:
  \eqn
\gamma_1(\pt) &=&  1 + 
 \frac{R^2}{2 \pi^2} p^2  \int \frac{d^2\kt}{(\kt^2)^2} d^2\kt_1 d^2\kt_2
\delta(\kt+\kt_1+\kt_2) \nn \\ 
&&\gamma_1(\kt_1)  \gamma_2(\kt_2)  
 \gamma_1(\kt+\pt) e^{i\kt_2(\bt_2-\bt_1)} \sin^2 \theta_{12}  \sin^2\theta_{p/k+p} \, ,  \nn \\
\gamma_2(\pt) &=&  1 + 
 \frac{R^2}{2 \pi^2} p^2  \int \frac{d^2\kt}{(\kt^2)^2} d^2\kt_1 d^2\kt_2
\delta(\kt+\kt_1+\kt_2) \nn \\ 
&&\gamma_1(\kt_1)  \gamma_2(\kt_2)  
 \gamma_2(\kt+\pt) e^{i\kt_1(\bt_1-\bt_2)} \sin^2 \theta_{12}  \sin^2\theta_{p/k+p} \, .
 \eqnx
 At least perturbatively, these equations imply the relations:
 \begin{equation}
 \gamma_2( \pt) = \gamma_1(-\pt) = \gamma_1^*(\pt)  ~~,~~ h(-\pt) = h^*(\pt)\, .
 \end{equation}
 Setting  finally  $\bt_2 = - \bt_1 = \bt/2$ we  get the basic integral equations:  
 \eqn
h(-\kt) &=& \frac{1}{\pi^2 \kt^2} \int d^2\kt_1 d^2\kt_2 \delta(\kt+\kt_1+\kt_2) e^{i(\kt_2-\kt_1)\bt/2} 
\gamma_1(\kt_1)  \gamma_1^*(\kt_2)   \sin^2 \theta_{12}
 \\
\gamma_1(\pt) &=&  1 + \frac{R^2}{2 } p^2  \int \frac{d^2 \kt}{\kt^2} h(-\kt) \gamma_1(\kt+ \pt) e^{-i\kt\bt/2} \sin^2\theta_{p/k+p}  \, ,   
 \eqnx
 which can be reduced to an integral equation for $\gamma_1(\pt)$ (or  $\gamma_2(\pt)$) alone:
 \eqn
\gamma_1(\pt) &=&  1 + 
 \frac{R^2}{2 \pi^2} p^2  \int \frac{d^2\kt}{(\kt^2)^2} d^2\kt_1 d^2\kt_2
\delta(\kt+\kt_1+\kt_2) \nn \\ 
&&\gamma_1(\kt_1)  \gamma_1^*(\kt_2)  
 \gamma_1(\kt+\pt) e^{i\kt_2\bt} \sin^2 \theta_{12}  \sin^2\theta_{p/k+p} \, .
 \label{gamma}
 \eqnx
 
 It is also quite easy to check that the action, on the e.o.m. takes the simpler form:
\begin{equation}
 A_{EOM} = \frac{Gs}{4\pi}  \int \frac{d^2\kt}{\kt^2} \left[3( e^{-i\kt \bt} \gamma_1(\kt) +  e^{i\kt \bt} \gamma_2(\kt)) - 2  e^{-i\kt \bt} \gamma_1(\kt) \gamma_2(-\kt)  \right]  \, .
  \label{OSA}
\end{equation}
An easy way to prove this is to note that our action, after elimination of $\phi$, is a functional of the $\gamma_i$ of the form:
\begin{equation}
 A  = A_1 + A_2 + A_4
\end{equation}
where $A_n$ is homogeneous in  $\gamma_i$  of degree $n$. By Euler's theorem:
\begin{equation}
 \int dk \gamma_i \frac{\delta A}{\delta \gamma_i}   = A_1 + 2 A_2 +4 A_4 \, .
\end{equation}
On the equations of motion this combination must be zero and therefore we can eliminate $A_4$ in terms of
$A_1$ and $A_2$:
\begin{equation}
 A_4  = -A_1/4 - A_2/2 \, ,
\end{equation}
 which gives the claimed result. 
This argument only works modulo boundary (surface) terms.  And, indeed, in the case discussed in the previous sections in which $\rho(0) \ne 0$, such boundary terms can be shown to be present since their absence would imply the relation:
\eq
I_{\phi} = - \frac12  [I_a + a(b^2)]\, ,
\eqx
which only holds if  $\rho(0)=0$. There are indications that the momentum-space approach automatically implies such a relation. Indeed, unless there is some singularity at small momenta, total derivative terms are set automatically to zero by momentum conservation. The boundary terms due to
$\rho(0) \ne 0$, on the other hand, come from a short-distance boundary which does not look to be present in the momentum approach.

In principle,  the above integral equations can be solved by iteration on a computer. The advantage, with respect to the position-space formulation, is that the iterative solution carries automatically with it   perturbative  boundary conditions. 
The iteration procedure is expected to converge only for sufficiently small values of $R/b$ where it should reproduce the perturbative expansion. 

Some preliminary  numerical  results have only been obtained \cite{NC} under the assumption that $\gamma(\kt)$ and $h(\kt)$ depend only on $\kt^2$ and $\bt^2$ but {\it not} on $\kt\cdot \bt$. This is only consistent with the field equations if we average them over the direction of $\bt$, i.e. if we make the replacement:
\begin{equation}
  e^{i\kt_2\bt} \rightarrow J_0(kb)\, ,
\end{equation}
which is the momentum space version of the  the azimuthal averaging procedure  made  in  sect. 4. 

A first interesting indication following from the numerical analysis \cite{NC}  is that, at sufficiently large $b$ where the iteration converges, $\gamma(k)$ approaches unity at small $k$ and a $b$-dependent constant larger than unity at large $k$.
This is fully consistent with the position space small-$r$ result:
\eq
\dot{a} \simeq - \frac{1}{ 2 \pi \dot{\rho}(b) r^2}  \, , \, \dot{\rho}(b) = \tanh \chi_b = 1- O(R^2/b^2)\, ,
\eqx
and confirms that, in the momentum-space approach,  one has automatically incorporated the condition $\rho(0)=0$.

 At some critical value of $R/b$ the iterative solution is found not  to converge any more,
  showing again the existence of a critical value for that ratio. One finds \cite{NC}
 $(b/R)_c \simeq 1.6 \pm 0.1$ a number that matches well (even too well!) our analytic estimate:  $(b/R)_c \simeq 2^{-1/2} 3^{3/4} \simeq 1.61$. A more accurate numerical calculation that does not use azimuthal averaging appears to give \cite{O} a slightly higher value, $(b/R)_c \simeq 2.38$. All these results are compatible with the CTS lower bound given in \cite{EG}, i.e. $(b/R)_c^{CTS} >  0.80$.
 
 We conclude that  the momentum-space approach gives numerical results for the boundary of the perturbative regime and the estimate of a critical $(b/R)_c$ that confirm those of the previous sections and are also compatible with the classical CTS-based collapse criteria.
We should stress, however,  that  numerical momentum-space techniques, being based on an iterative procedure, cannot be easily extended  below $b = b_c$. Searching for complex solutions remains a serious challenge.

\subsection{Arguments for the existence of a critical $b/R$ ratio }

A quick -- though approximate -- way to argue for the existence of a critical $R/b$ comes by considering the integral equation (\ref{gamma})  for $p \sim 1/R < 1/b$ and to realize that, in this case, the integrals should be dominated by the regions in which all the arguments are roughly of the same order $\sim 1/R$.
The equation for  this ``average" $\bar{\gamma}$ then takes the form:
\begin{equation}
\label{average}
  \bar{\gamma} = 1 + K^2 (R/b)^2  \bar{\gamma}^3\, ,
\end{equation}
where $K$ is a numerical constant of $O(1)$. It is easy to see that the perturbative solution $\bar{\gamma} = 1 + \dots$  ceases to exist above a critical value of  $R/b$, $(R/b)_c = \frac{2}{3\sqrt{3}K}$.
Near this critical point the action becomes singular with a $(b-b_c)^{3/2}$ behaviour similar to the one found in sect. 4.

It is perhaps worthwhile to notice at this point an intriguing relation between  eq. (\ref{average}) and the equation determining the turning point $r=r^*$ for a null geodesic impinging on a Schwarzschild metric of radius $R$ at impact parameter $b$. In this latter case the equation reads:
\begin{equation}
\label{turning}
 \frac{R}{b} = x - x^3 ~~, ~~ x \equiv \frac{r^*}{b}\, ,
\end{equation}
giving the well-known result that the turning point disappears for $b < b_c = \frac{3\sqrt{3}}{2}R$. Similarly, eq. (\ref{average})  can be put in the form:

\begin{equation}
\label{ave}
 K \frac{R}{b}  = y - y^3 ~~, ~~ y  \equiv K (R/b)  \bar{\gamma}\, ,
\end{equation}
giving, for $K=1$,  the same critical value for $b/R$.

Amusingly, eq. (\ref{critical}) takes a similar form, this time in terms of $R^2/b^2$, i.e.
\begin{equation}
\label{ave}
 \frac{R^2}{b^2}  = z - z^3 ~~, ~~ z  \equiv \dot{\rho}(b^2)  \, ,
\end{equation}
and thus gives again the critical value $\frac{3\sqrt{3}}{2}$, though  for $R^2/b^2$.

The above reasoning suggests that a more rigorous argument for the existence of a critical $b/R$ could possibly be constructed along the following lines.
In eq. (\ref{gamma}) rescale all momenta by a factor $|b|$ and distinguish the new dimensionless momenta from the old ones by a tilde. Also, multiply both sides of the equation by a factor $R/b$ and define $\beta = (R/b) \gamma_1$: Then eq. (\ref{gamma}) takes the form:
 \eqn
\frac{R}{b}  &=&  \beta(\tilde{\pt}) -
 \frac{\tilde{\pt}^2 }{2 \pi^2}  \int \frac{d^2 \tilde{\kt}}{(\tilde{\kt}^2)^2} d^2\tilde{\kt}_1 d^2\tilde{\kt}_2
\delta(\tilde{\kt}+\tilde{\kt}_1+\tilde{\kt}_2) \nn \\ 
&&\beta(\tilde{\kt}_1)  \beta^*(\tilde{\kt}_2)  
 \beta(\tilde{\kt}+\tilde{\pt}) e^{i \tilde{\kt}_2 \boldsymbol{e}} \sin^2 \theta_{12}  \sin^2\theta_{p/q+p} \, ,
 \label{beta}
 \eqnx
where $\boldsymbol{e}$ is the unit vector in the direction of $\bt$.

The rhs of the new equation is a functional of $\beta$ and a function of $\tilde{\pt}$. When the lhs $R/b$
is very small the equation can be solved by taking $\beta$ small so that the cubic term on the rhs is negligible. But when $R/b$ is sufficiently large this is no longer the case. Taking $\beta$ large and of order $R/b$ may not help to find a solution if the cubic term takes over and has the wrong sign.
Hence it should not be impossible to show, by some functional analysis, that this equation does not have perturbative solutions for sufficiently large values of $R/b$. We do not attempt such a proof here.

\section{Particle production and inelastic unitarity}
\setcounter{equation}{0}

So far we have neglected the imaginary part of $\delta$, i.e. the phenomena associated with the production of gravitons. Even when we limit ourselves to the IR-safe (TT) polarization there are at least two interesting issues to be addressed: one concerns the spectrum of the produced gravitons (as a function of their transverse momentum); the other is the damping of the elastic amplitude caused by
the opening of inelastic channels. We would like to study both effects as a function of $b$ (or actually $b/R$) in order to see whether some interesting physics shows up as we approach a critical value.

A convenient way to study production amplitudes within our effective action approach is to introduce an auxiliary source $J$ coupled to the field $h$ that corresponds to the physical $TT$ graviton. 
At the same time, in order to ensure full inelastic unitarity, contributions to the action from ${\rm Im}~ \delta$ have to be included. 
The standard procedure would be to  do all this at the level  of the 4-dimensional action (\ref{action}) by coupling the source $J$ to the canonically normalized 4-D field $\frac14 \nabla^2 \Phi$.
We shall instead use a ``shortcut" and modify directly the reduced action (\ref{reduced_a})
as follows:
\eqn
\label{modA}
\frac{\pi \A}{Gs} \rightarrow \frac{\pi \tilde{\A}}{Gs} = \frac{\pi \A}{Gs} + 
\frac{(\pi R)^2}{2} \int d^2 \kt \left(  \frac{1}{2} h(\kt)  h(-\kt) \frac{2iY}{\pi}  + h(-\kt)  \frac{2  \sqrt{Y}}{\pi R \sqrt{Gs}} J(\kt) \right) \, ,
\eqnx
where the explicit  $Y = \log s$ dependence takes effectively into account longitudinal phase space.
We will also interpret the additional terms in the action as being defined ``on-shell" i.e. on the equations of motion of the unperturbed action. This is  correct as far as the additional source term is concerned,  but probably an oversimplification for the additional imaginary part in (\ref{modA}) which presumably
changes the field equations. With this caveat we shall now proceed to the computation of the full (inelastic) $S$-matrix.

The  functional:
\eqn
e^{iW(J)} = e^{i \tilde{\A}_{eom}(J)} \, ,
\eqnx
will generate, through its functional derivatives with respect to $J$, the scattering amplitudes for producing an arbitrary number of gravitons:
\eqn
S(2\rightarrow 2 + \kt_1+\kt_2+\dots \kt_n) = \left( \frac{\delta^n}{\delta J(-\kt_1)\dots\delta J(-\kt_n)} e^{iW(J)}\right)_{J=0} \, . 
\eqnx

This prescription  can be checked to reproduce, at lowest order, the one-TT-graviton production cross section --and hence the imaginary part of the elastic scattering amplitude--  as  given both by the H-diagram. 
In order to consider the general multi-graviton production amplitude we note  that, at sufficiently high energies, a WKB-like approximation holds (since, at very high energies, $\delta^2  W/ (\delta J)^2 \ll (\delta W/ \delta J)^2$, etc)  :
\eqn
&&  \left( \frac{\delta^n}{\delta J(-\kt_1)\dots\delta J(-\kt_n)} e^{iW(J)}\right)_{J=0}  
 =  \nn \\
&&  \left(e^{iW(J)}\right)_{J=0} \left(\frac{i \delta  W}{\delta J(\kt_1)}\right)_{J=0}   \dots \left(\frac{i \delta W}{\delta J(\kt_n)}\right)_{J=0} +{\rm  subleading ~ terms}\, .
\eqnx

As a consequence, the  emitted gravitons are approximately uncorrelated and the multi-graviton amplitude factorizes. Furthermore since, on the equations of motion,
\eqn
 \left(\frac{i \delta  \tilde{\A}}{\delta J(-\kt)}\right)_{J=0}  = i \sqrt{ Gs Y} R ~ h(\kt)_{J=0}\, ,
\eqnx
we find:
\eqn
S(2\rightarrow 2 + \kt_1+\kt_2+\dots \kt_n) =  \left(e^{iW(J)}\right)_{J=0} (i \sqrt{Gs Y} R)^n
   \prod_i   h(\kt_i)_{J=0}\, ,
\eqnx
and 
\eqn
\frac{1}{\sigma_{el}}  \frac{d \sigma (2\rightarrow 2 +  \kt_1+\kt_2+\dots \kt_n)}{ d^2\kt_1 \dots d^2\kt_n} =  
   (Gs Y R^2)^n  \prod_i   |h(\kt_i)_{J=0}|^2 \, ,
\eqnx
with $h(\kt)$ given in eq. (\ref{h=H}).
It is easy to double check that, at the lowest level and for $n=1$,  this reproduces the one-graviton cross section discussed in sect. 2.

At the same time, the elastic amplitude will be absorbed. Its absolute square will be controlled by the
imaginary part of the on-shell action at $J=0$ and can be easily computed again in terms of $h(k)$.
We find:
\eqn\label{sel}
\sigma_{el} =  |S(2\rightarrow 2 )|^2 = \exp (-2 Im \tilde{\A}_{J=0}) = \exp\left( - Gs R^2 Y \int d^2 \kt  |h(\kt)|^2 \right)  \, ,
\eqnx
i.e. precisely in such a way as to ensure the inelastic unitarity of the S-matrix.

These results can be summarized  by writing the S-matrix in an operator form involving also the longitudinal-momentum  degrees of freedom:
\eqn \label{OSM}
S &=& \exp \left(i \sqrt{Gs} R \int \frac{d^3 k}{\sqrt{k_0}} (h(\kt) \hat{a}_k + h(\kt)^* \hat{a}^{\dagger}_k) \right) =  \exp\left( - Gs R^2 \int \frac{d^3 k}{2 k_0}  |h(\kt)|^2 \right)  \nn \\
&& \exp \left(i \sqrt{Gs} R \int \frac{d^3 k}{\sqrt{k_0}} h(\kt)^* \hat{a}^{\dagger}_k \right)  \exp \left( i\sqrt{Gs} R \int \frac{d^3 k}{\sqrt{k_0}} h(\kt) \hat{a}_k \right)\, ,
\eqnx
where $\hat{a}_k $ , $\hat{a}^{\dagger}_k$ are canonically-normalized destruction and creation operators of physical gravitons of momentum $k$ and IR-safe polarization and we have left out for simplicity an overall $c$-number phase $\exp{i\A(b, s)}$ containing the on-shell uncorrected action. This gives back, for instance, eq. (\ref{sel})
after realizing  that the longitudinal momentum integration just provides a factor $Y= \log s$.

The $S$-matrix (\ref{OSM}) when acting on the Fock vacuum of the $\hat{a}_k $ , $\hat{a}^{\dagger}_k$ operators creates a coherent state of physical gravitons in which we can compute the expectation value of the associated canonical quantum field that we denote by $h_{\rm can}$. This is best done by using the LSZ formalism according to which:
\eqn\label{LSZ}
S(2\rightarrow 2 +k) = \sqrt{2 k_0} \langle 2 | a^{{\rm out}} | 2 \rangle = 
i \int \frac{d^4 x}{(2 \pi)^{3/2}}  e^{ikx}  \partial_{\mu}  \partial^{\mu}   \langle 2 | h_{{\rm can}}(x) | 2 \rangle|_{k^2\rightarrow 0}\, ,
\eqnx
where, in our case, the canonical $TT$-graviton field is given by:
\eqn\label{can}
h_{\rm {can}} = (8 \pi G)^{-1/2} \frac{\epsilon^{\mu\nu}_{TT} }{\sqrt2} h_{\mu\nu}(x)\, ,
\eqnx 
and the extra factor $1/\sqrt{2}$ comes from our normalization of the polarization tensors.

Using properties of the coherent state  generated by (\ref{OSM}) it is easy to check that the following metric fluctuation satisfies (\ref{LSZ}):

\eqn\label{metricCS}
\langle h(x)_{\mu\nu} \rangle = - \frac{ R^2}{4 \pi^2}  \int \frac{d^4 k }{k^2 + i \epsilon}  \epsilon_{\mu\nu}^{TT} 
 h(\kt) e^{-ikx} \, ,
\eqnx 
 where, as previously explained, $\epsilon_{\mu\nu}^{TT} $ act as differential operators and we have inserted an $i \epsilon$ prescription although the LSZ formula is only sensitive to the principal part of the propagator.

Equation (\ref{metricCS})  should  be consistent with the effective metric  of eq. (\ref{metric_tot}) and, at lowest order in $R/b$, with the one that follows from  eq. (\ref{Hbypol}).  Indeed, if one looks at the principal part contribution, one finds, in the small-$\kt$ limit \footnote{Keeping the $\kt$-dependence amounts to multiplying the $\Theta$-function by $J_0(|\kt|\sqrt{x^+x^-})$, which implies the cutoff $x^+x^-\lesssim R^2$ mentioned in sec.2.3.} implicit in our procedure:
\eqn \label{metric CS2}
\langle h(x)_{\mu\nu} \rangle =  ( \pi R)^2  \int d[\kt] \epsilon_{\mu\nu}^{TT} 
 h(\kt) \Theta(x^+ x^-) e^{-i\kt \xt} =  ( \pi R)^2  \epsilon_{\mu\nu}^{TT} 
  \Theta(x^+ x^-) \nabla^2 \phi \, ,
\eqnx
which explains the normalization of the $\phi$ field used in eq. (\ref{metric_tot}).

Using the small-$k$ limit of $h(k)$ we find $h_{cl}(x) \sim R^2/r^2$ at large $r$, in agreement with (\ref{Hbypol}) but in apparent disagreement  with the standard quadrupole formula, which would require $h_{cl}(x) \sim R^2/b r$.  This can possibly be explained by the fact that the TT polarization is emitted mainly in the forward and backward direction $|k| \gg |\kt|$ so that there is still a non-trivial flux of TT-gravitons at null infinity. On the other hand,  the  (IR-unsafe) polarization that we neglected  exibits a $1/r$ behaviour in agreement with the fact that IR singularities are associated with classical radiation.

It is very tempting, at this point,  to guess a generalization of our result (\ref{OSM}) to include the IR-sensitive polarization. This would read:
\eqn \label{OSM2}
S = \exp \left(\sqrt{Gs} R \int \frac{d^3 k}{\sqrt{k_0}} \left(i h_{TT} (\kt) \hat{a}_k - h_{LT} (\kt) \hat{b}_k - h.c. \right) \right) \, ,
\eqnx
 where $h_{TT} = h$ and:
 \eqn
 h_{LT}  =  \frac{1}{\pi^2 k^2} \int d^2\kt_1 d^2\kt_2 \delta(\kt+\kt_1+\kt_2) e^{i(\kt_2-\kt_1)\bt/2} 
\gamma_1(\kt_1)  \gamma_2 (\kt_2)   \sin \theta_{12}  \cos \theta_{12}\, .
 \eqnx
 
This would provide a unitary S-matrix whose matrix elements, however,  are only finite within particular coherent states that include soft bremmstrahlung, a well-know situation in perturbative QED and quantum gravity, already discussed in \cite{ACV2}. Once this is properly done there should be no major conceptual obstacle in including the effects of the LT polarization on the gravitational collapse problem.

What remains to be done is to evaluate $h(k)$ in different regimes  in order  to extract both the spectrum of the emitted gravitons and the absorption of the elastic amplitude as a function of  $b/R$. This can only be done, of course, after solving the classical equations, similarly to what already done for $\rm{Re} \A$ in position space.
We can summarize our present understanding on this matter as follows:
\begin{itemize}
\item At $b \gg  R$ the spectra have a logarithmic behaviour at small $k$ and an exponential damping at $k \gg   b^{-1}$. In other words the typical tranverse momenta of the produced gravitons are, not surprisingly, of the same order as those of the exchanged gravitons (see the discussion in sec. 2). In turn, the elastic amplitude is suppressed by $\exp{(-GsY R^2/b^2)}$. Note that, for $R > l_s$,  such a suppression dominates over the one   due to string excitation, that we have neglected.
\item As one approaches the region $b \rightarrow b_c \sim R$ from above, physics appears to be rather smooth. Nevertheless, the graviton spectrum is now cutoff at momenta of order $1/R$ (i.e. of order of the Hawking temperature of a BH of mass $\sqrt{s}$) and the elastic amplitude is suppressed by an exponential factor  $\exp{(- c GsY )}$ (with c some constant of O(1)) which, modulo the factor $Y = \log s$ corresponds to $\exp{(-{\cal S})}$, with ${\cal S}$ the Bekenstein-Hawking BH entropy
\item It would be very interesting to find out what happens if one goes to the  region $b << R$, in particular whether the cut-off on momenta keeps growing like $1/b$ or remains ``frozen" at $1/R$ as black-hole evaporation would suggest.
It is not yet fully clear how that region can be studied. The possibility of an analytic continuation of S-matrix elements and effective action solutions for $b<b_c$ is discussed in the next Section.
\end{itemize}

\section{Action and complex solutions for $b<b_c$}
\setcounter{equation}{0}

We have realized in sec. \ref{s:$b>0$} that real-valued field solutions with $\rho(0)=0$ exist only for $b>b_c$, and that, below $b_c$, they become complex. We have also shown that the action has a
branch-point singularity at $b=b_c$, presumably due to the pinch of two such solutions of the criticality equation (\ref{critical}). The problem then arises of how to define both S-matrix and solutions for $b<b_c$.

It is tempting to try the simplest possibility, and to continue analytically the S-matrix on the basis of the form $\sim (b^2-6\sqrt{3}G^2s)^{3/2}$ of the branch-cut, by choosing a physical energy-sheet reached by an $s+i\epsilon$ prescription. According to the expansion (\ref{expansion_a}) this one corresponds to the $\chi_b-\chi_c$ determination having a negative imaginary part and contributes a positive imaginary part to the action. This criterion tells us to take that particular complex solution as the physical one on which the action, and thus the S-matrix, should be computed.

An additional argument for this choice comes from the tentative interpretation of the $\rho(0)=0$ condition as stationarity equation of the integral over $\chi_b$ in eq.~(\ref{chi_integral}). If we define the analytic continuation of the action by that integral -- which, for $b<b_c$ has two complex conjugate stationarity points -- we should take the one for which the saddle point is stable. Then, by ignoring the measure factor, calling $\chi_b$ the stationarity point and simply $\chi$ the integration variable, we have the expansion
\eqn\label{expansion_b}
\frac{\A(b,s;\chi)-\A(b,s;\chi_b)}{Gs}~&=&~2(-1+\frac{b^2\sinh \chi_b}{R^2\cosh^3\chi_b})(\chi-\chi_b)\\ \nn ~&+&~\frac{b^2}{R^2}(1-t_b^2)(1-3t_b^2)(\chi-\chi_b)^2+ O((\chi-\chi_b)^3)\, .
\eqnx 
We note again that the action is stationary on the solutions of (\ref{critical}) and that the fluctuation coefficient in eq.~(\ref{expansion_b}) becomes, for small $b-b_c$,
\eq\label{fluctua}
\frac{1}{t_b}(1-3t_b^2)\simeq \mp 2 \sqrt{2} i\sqrt{1-\frac{b^2}{b_c^2}},~~~(t_b-t_c)\simeq\pm i\frac{\sqrt{2}}{3}\sqrt{1-\frac{b^2}{b_c^2}}\, .
\eqx
The saddle point is then stable when the above coefficient is positive imaginary, corresponding to damped fluctuations, yielding again the solution with Im$t_b<0$ for which the action acquires a positive imaginary part, as noticed before. 

By then taking for $b\lesssim b_c$ the complex solution for $t_b$ or $\chi_b$ with negative imaginary part, we obtain, from eq.~(\ref{expansion_a}),
\eq
\label{elatbc}
 \frac{{\rm Im}\A(b,s)}{Gs}~=~\frac{2\sqrt{2}}{3}(1-\frac{b^2}{b_c^2})^{3/2} \, , |S(b,s|^2\simeq 
 \exp\left( - \frac{2\sqrt{2}}{3} \frac{R\sqrt{s}}{\hbar} (1-\frac{b^2}{b_c^2})^{3/2} \right)\, .
\eqx
Equation (\ref{elatbc}) implies that the S-matrix, when analytically  continued to $b<b_c$, acquires an additional absorptive part (on top of the one due to TT-graviton  production discussed in the previous section) whose interpretation calls for the opening up of some extra channels in this new regime. It is tempting to think of these as quantum analogs of the black-holes that are expected to be formed on glassical grounds \cite{EG}. At the same time, the dominant solution is also complex-valued, with Im$\rho(r^2)\le 0$ for $r\simeq R$. 
 
The above features are confirmed by evolving the complex solution to smaller values of $b$. In the $b \rightarrow 0$ limit the stable determination becomes, by eqs.~(\ref{critical}) and (\ref{fluctua}),
\eq\label{smallb}
t_b \simeq e^{-i\pi /3}(\frac{R}{b})^{2/3},~~~\chi_b=-i\frac{\pi}{2}+e^{i\pi /3}(\frac{b}{R})^{2/3},~~~(b\ll R)\, ,
\eqx
while the complex conjugate solution is unstable. Correspondingly, the action for $b\ll R$ takes the form
\eq\label{elat0}
\A(b,s)~=~Gs(2\chi(L^2)+i\pi-3e^{i\pi/3}(\frac{b}{R})^{2/3}),~~~|S_{el} (b,s|^2\simeq e^{- R\sqrt{s}(\pi - \sqrt{6} (b/R)^{2/3})}\, . 
\eqx
For $b=0$ the suppression factor in the elastic cross section is just 
$\exp(-{\cal S}_{\rm BH}(\sqrt{s/2}))$, with $\cal S_{{\rm BH}} ({\rm M})$ the Bekenstein-Hawking entropy of a Schwarzschild black hole of mass $M$. This is compatible with a statistical interpretation where a fraction $1/\sqrt{2}$ of the incoming energy goes into forming a black hole, even if the
relationship of our S-matrix framework to such a semi-classical statistical picture is yet to be
clarified.

On the other hand, the suppression (\ref{elatbc}) of the elastic amplitude  appears to die off as $(1-\frac{b^2}{b_c^2})^{3/2}$ for $ b \rightarrow b_c^-$. This looks like an interesting (and we believe robust) result calling for a physical interpretation. From a classical standpoint, this limit should correspond to the production of a nearly extremal Kerr black hole with $J \le J_c = G M^2$. If the mass of the produced black hole would  remain finite in this limit its entropy would approach a finite value (just half of that of a Schwarzschild black hole of the same mass) and the statistical interpretation invoked earlier for 
$b \sim 0$ would fail. 
However, it is  conceivable (although, to the best of our knowledge, still not proven) that in a nearly critical-collapse situation most of the initial energy and angular momentum are radiated away to infinity leaving only a vanishing
 mass and angular momentum to collapse  at the critical point.
In this case, if we insist on identifying our elastic suppression with an entropy factor, we  have to assume that the mass of the Kerr  black hole being formed in the collision vanishes like $(b_c-b)^{3/4}$  for $b \rightarrow b_c^-$. This would imply a ``Choptuik exponent" of $0.75$ in our critical collapse, i.e. about twice the original exponent of $\sim 0.37$  found in Choptuik's original paper \cite{Chop}.\footnote{One of us (G.V.) would like to thank  Steve Giddings and Don Marolf  for interesting discussions about this possibility.}

Corresponding to the small-$b$ parameters (\ref{smallb}), the expression for the physical field solution $\rho(r^2)$ has the initial value
\eq\label{smallb_sol}
\frac{\rho(b^2)}{R^2}~=~\frac{b^2}{R^2}t_b~=~e^{-i\pi /3}(\frac{b}{R})^{4/3}\, ,
\eqx
which is consistent with $\rho(0)=0$ and has a small, positive real part also. This is sufficient to have an $r^2$-evolution of $\rho(r^2)$ with increasing real part and negative imaginary part, which tunnels to a perturbative, real valued  behaviour at large distances. We can further check that the action integrals in (\ref{IaIphi}) are well defined on the physical solution, and their evaluation holds unchanged, except that $\rho(0)=0$ is now built in, so that the result (\ref{actionb1}), and thus (\ref{elatbc}) and (\ref{elat0}) obtain automatically, with the appropriate (complex) values of $\chi_b$ and $t_b$.

 In order to better understand the $r^2$-evolution, it is convenient to come back to the axisymmetric $b=0$ case, by looking for possibly complex solutions satisfying the boundary condition $\rho(0)=0$. Since the parametric expressions in eqs.~(\ref{soln_a}) and (\ref{soln_b}) are still valid, we obtain $\rho(0)=0$ by setting $\chi_0=-i\pi/2$. With this boundary value, for $r\ll R$ it is convenient to look for solutions $\chi(r^2)\equiv -i\pi/2+\eta(r^2)$ such that, by (\ref{soln_a}),
\eq\label{complexp}
2\frac{r^2}{R^2}=2\chi+\sinh2\chi+i\pi~=~2\eta-\sinh2\eta\simeq-\frac{4}{3}\eta^3+O(\eta^5)\, .
\eqx
This equation admits in turn three branches, according to the values $\epsilon=(e^{\pm i\pi/3},-1)$  of the three cubic roots of $(-1)$, as follows
\eqn\label{complfields}
\eta(r^2)~&\simeq&~\epsilon~(\frac{3r^2}{2R^2})^{\frac{1}{3}}, ~~~\frac{\rho(r^2)}{R^2}~\simeq~-\eta^2~\simeq~\epsilon^*(~\frac{3r^2}{R^2})^{\frac{2}{3}},~~(r\ll R)\\ \nn
R^2\dot{a}&\simeq&-\frac{1}{2\pi}\epsilon(\frac{3r^2}{2R^2})^{-2/3},~~~(2\pi r)^2\dot{\phi}~\simeq~-\frac{\rho}{R^2}\simeq -\epsilon^*(~\frac{3r^2}{2R^2})^{\frac{2}{3}}\, .
\eqnx
We thus see that the solution with $\epsilon=e^{i\pi/3}$ matches the physical one discussed above in the $b\rightarrow 0$ limit. We also see that the small-$r^2$ exponents are fractional, and the delta-function flux from $r=0$ in (\ref{eom_1}) is shared between $\dot{a}$ and $\dot{\phi}$. Note that there is also a real-valued solution (the one with $\rho\le 0$ noticed before), which however is unphysical: it is quite nonperturbative ($\dot{\rho}=-1$ in the large-$r$ limit) and yields a quadratic IR divergence in the action. Note finally that the large-$r^2$ behaviour of the physical fields is instead perturbative, the value of $\chi_0=-i\pi/2$ contributing (via the scale $\bar{r}^2$ in (\ref{asy_a})) some subleading imaginary part.

For $b>0$, we note that the small-$r$ behaviour of $\phi$ changes in the $r<b$ region. In fact, since in this region $\ddot{\rho}=0$, then $\dot{\rho}=t_b$ must be a constant and, due to $\rho(0)=0$, $\dot{\phi}$ will be a constant too:
\eq
(2\pi R)^2\dot{\phi}=1-t_b=1-\epsilon^*~(\frac{b^2}{R^2})^{-1/3};~~~\ddot{\phi}=0~~~(r<b)\, .
\eqx
Therefore, $b>0$ acts effectively as a cutoff for the small-distance behavior $\dot{\phi}\sim d(r^2\dot{\phi})/dr^2\sim r^{-2/3}$ occurring at $b=0$. The $\dot{\phi}$ field becomes completely regular.

We thus see that the small-$r$ behaviour of the complex physical solution has now changed -- compared to the real-valued ones -- to a ``weak-field'' profile\footnote{We use inverted commas here since the perturbation of the metric is actually still large at $r \sim R$ and even more so at $r \sim b$.}, due essentially to the $\rho(0)=0$ property. The latter condition acts in this context as a quantization rule, yielding a well-defined solution whose classical counterpart, if any, one should classify as being  ``untrapped''. Therefore, at quantum level, the physical solutions for $b<b_c$ show no evidence of a field being confined behind the would-be horizon.

 The main challenge remains, of course, the interpretation of such complex solutions and of the extra absorption found in the elastic channel. A possible way to proceed is to continue below $b_c$ the production amplitudes of the multi-particle channels considered so far, and look for unitarity integrals that could make up for the extra absorption.  Since the field $h\sim 1-\dot{\rho}$ acquires an imaginary part, the latter could perhaps be interpreted according to the $h$-definition in sec. (2.1): the imaginary part would then simply be the $LT$ polarization which, though excluded in the beginning, is turned on necessarily below $b_c$. In any case, within this interpretation, the Hawking evaporation required by the additional absorptive part should be looked for in the various contributions of such imaginary part to the multi-graviton spectra. Since there is no $\sim \log s$ longitudinal phase space enhancement, it should be quite central, with  $k_3\sim |\kt|\sim 1/R$ and  an emission yield $\sim {\rm Im} \A$.
  
 Another possible  interpretation (see sect. 7)  is that the missing probability goes into the formation
 of some new  bound states. One may object that such bound states, being very massive, should decay into light particles (gravitons in our case), behave as resonances, and that consequently the $S$-matrix should already be unitary in the multigraviton  Hilbert space. However, it is quite possible that, within our approximation that neglects corrections of relative order $\hbar/Gs \equiv M_P^2/s  \ll 1$, such resonances are actually stable and have to be included in the possible final states in order to ensure unitarity.  For instance, according to standard lore, black-holes  have a lifetime of order $R$ and thus  propagators of the type $(s-M_{BH}^2+i M_P^2)^{-1}$. If objects of this kind, even if not necessarily to be identified as black holes, were responsible for the extra absorption below $b_c$,   their  finite-width  effects would be lost  in our semiclassical regime, $Gs/\hbar \gg 1$.

\section{Summary and outlook}
\setcounter{equation}{0}

Let us summarize the method we have used, and the main assumptions and results of our investigation.

Working in the superstring approach to scattering amplitudes in the transplanckian regime $Gs\gg \hbar$~\cite{ACV1, ACV4, ACV2}, we have used the effective action framework justified in~\cite{ACV3} for the case $R,b\gg\ls$, in which string-size effects are not very important. We have then neglected the so-called rescattering terms (as it should be justified for fixed (not too small) $b/R$) and, in order to avoid the known~\cite{ACV2} but technical treatment of the IR problem, we have considered the emission of only one graviton polarization, the IR-safe one, described here by the scalar field $\phi$. Both are technical simplifications that could in principle be waived one by one.

In the framework just described (sec.2), the dependence of the effective action and fields on the longitudinal coordinates simplifies, so that the problem reduces to a transverse two-dimensional effective action, that we have considered in both configuration (secs.3, 4) and momentum space (secs.5, 6). In momentum space, an iterative procedure for solving the equations of motion has been set up, and is suitable for numerical computations. In configuration space, the equations are studied in the axisymmetric case, in which they reduce to ordinary differential equations. Of course, axisymmetric solutions are directly relevant only in the $b=0$ limit, while for $b>0$ they imply the azimuthal averaging procedure explained in sec. 4.  
 
A key point we have discussed throughout the paper is about the boundary conditions to be set in order to determine both field solutions and action. One is provided by matching the perturbative behaviour for $r\gg b, R$. The other ($\rho(0)=0$) is expressed in terms of the auxiliary field $\rho=r^2(1-(2\pi R)^2\dot{\phi})$ and is still suggested by the weak coupling regime valid for $b\gg R$. In addition, we argue that it has to be valid in the non-perturbative regime also, as the only consistent way to treat the $r=0$ boundary. Under such conditions, we are able to provide analytic field solutions, the corresponding action and effective metric and then the phaseshift operator resumming the $R^2/b^2$ corrections to the eikonal and the S-matrix. 
 
The perturbative resummation diverges at a critical value of the impact parameter $b=b_c\sim R=2G\sqrt{s}$, which separates the class of real-valued ($b>b_c$, sec.4) and complex-valued ($b<b_c$, sec.7)  solutions, all satisfying the boundary condition $\rho(0)=0$ which plays the role of quantization condition of the problem. We also find that our estimates of the $b_c/R$ ratio are compatible with the classical lower bound for CTS formation, suggesting that our non-perturbative regime is likely to be in correspondence to classical collapse.   

For $b>b_c$, the S-matrix has essentially the form of a unitary coherent state operator from which elastic absorption and inelastic spectra can be computed. For $b\simeq b_c$, the graviton spectrum is cutoff at transverse momenta of order $\hbar/R=T_H$, the Hawking temperature of a black hole of mass $\sqrt{s}$, and the corresponding elastic absorption is $\sim\exp(-$const.$~GsY)$. 

For $b<b_c$, the analytically continued physical field solution with $\rho(0)=0$ is complex, and yields an additional elastic absorption compared to that just mentioned. The absorptive suppression is exponential, and the exponent vanishes like $Gs (1-b^2/b_c^2)^{3/2}$ for $b\rightarrow b_c$ and, for $b=0$, is just $2\pi Gs$.  This behaviour is compatible, as order of magnitude, with
the entropy of a black hole of properly chosen mass, as argued in sec. 7, even
if the relationship of our S-matrix coherent state to such a classical object  is
yet to be clarified.

Our field solutions provide an effective metric also, which is of shock wave type for the longitudinal fields, and of finite wavefront for the mostly transverse one. For $b>b_c$, the profile function of the transverse field $\phi$ is everywhere regular in the transverse $r$ coordinate. For $b<b_c$ some metric components become complex and their interpretation is open to discussion but, surprisingly, the $\phi$ field keeps being regular for $0<b<b_c$ also, and has a mild fractional $r=0$ singularity for $b=0$. We refer to this feature as a ``weak field'' situation (see however our footnote in sect. 7). In this sense, our physical solutions, for any value of $b$, show no evidence of a field being confined in the small-$r$ region. 

 For $b<b_c$, we have studied real valued field solutions also, which exist at the expense of violating the boundary condition $\rho(0)=0$. They have in fact a positive $\rho(0)$ and thus show a $\dot{\phi}\simeq-\rho(0)/r^2$ singularity and a ``strong field'' situation. Such solutions are however ill-defined (depending on the value of an arbitrary parameter), just because the $\rho(0)=0$ condition is not met. For this reason, we believe them to be unphysical. Then, if the above ``quantization'' condition is imposed, the solutions for $b<b_c$ become complex and change to the ``weak field'' profile discussed before. The correct solution is therefore no longer confined, in a way suggestive of a quantum tunnel effect.

The additional absorption found before for $b<b_c$ calls for extra production channels for the S-matrix to be unitary, on whose nature we have made a couple of guesses in sec. 7. Here, further work is needed in order to continue the appropriate production amplitudes below $b_c$ and thus to check whether inelastic unitarity is really verified: in this respect, our results are only partial. Nevertheless, we feel that some new physics is emerging in this simplified, but consistent quantum-gravity treatment that we have proposed. The picture outlined by our results suggests that this is the right framework for at least asking the questions, even if we only have some of the answers. 

\section*{Acknowledgements}
D. A. thanks Stefano
Liberati for stimulating discussions on regular BH solutions in
semiclassical gravity.
M. C. wishes to thank Dimitri Colferai for interesting discussions and the CERN Theory Division for hospitality while part of this work was being done.
G. V. thanks  Giuseppe  Marchesini, Enrico Onofri and Jacek  Wosiek for  discussions and for communicating  to him the numerical results reported here. He also acknowledges discussions with David Gross, Steve Giddings and Don Marolf as well as the hospitality of the Department of Physics and Astronomy at UCLA during the final stages of this work.
The authors have enjoyed the hospitality of the Galileo Galilei Institute in Florence during various phases of their collaboration.
This  work was supported in part by a PRIN grant (MIUR, Italy) and by the EU grant
MRTN/CT/2004/503369.
 
\section*{Appendix: Boundary terms and the $\rho(0)=0$ condition}
\renewcommand{\theequation}{\rm{A}.\arabic{equation}}
\setcounter{equation}{0}
Here we investigate the boundary terms present in the definition (\ref{reduced_a}) with the purpose of understanding in a more formal way the boundary condition $\rho(0)=0$. The expression (\ref{reduced_a}) is nominally of fourth order in the derivatives of $\phi$, but can be reduced to second order by introducing as fundamental field a first derivative. Furthermore, the expression depends on the current $\Hc$ which, for generic fields $a$ and $\bar{a}$, is nonlocal. We shall thus look for a local form of the effective lagrangian, by specializing  for simplicity to the axisymmetric case, in which all fields are functions of $r^2$ only. 

We thus express the $\dot{\phi}$ field in eq.~(\ref{reduced_a}) in terms of the $\rho$ function introduced in eq.(\ref{axibconst}),namely
\eq
\rho(r^2)~\equiv~r^2(1-(2\pi R)^2\dot{\phi}(r^2))\, ,
\eqx
 and we solve for $\Hc(r^2)$ in terms of $\dot{a}\dot{\bar{a}}$ by the same manipulations of secs. 3 and 4. We then obtain
\eqn\label{boundary_a}
\frac{\A}{2\pi Gs}~&=&~a(b^2)+\bar{a}(0)+2\pi\int_0^{\infty}dr^2\left(-r^2\dot{a}\dot{\bar{a}}+2\Hc (r^2)(1-\dot{\rho})-\frac{(1-\dot{\rho})^2}{(2\pi R)^2}\right);\\ \label{boundary_b} 2 \Hc (r^2)~&=&~\int_{r^2}^{\infty}dr^2\dot{a}(r^2)\dot{\bar{a}}(r^2)\, ,
\eqnx 
where we have specialized to the determination of $\Hc$ which vanishes at large distances -- similarly to what we have done for $\dot{a}$ on the equations of motion -- in order to recover the perturbative behaviour in that region.
 We then perform an integration by parts in order to eliminate the $r^2$-integral in the expression of $\Hc$
\eq\label{byparts}
\int_0^{\infty}dr^2~2\Hc (r^2)(1-\dot{\rho})-\dot{a}(r^2)\dot{\bar{a}}(r^2)(r^2-\rho(r^2))~=~\rho(0)\int_0^{\infty}dr^2\dot{a}(r^2)\dot{\bar{a}}(r^2)\, ,
\eqx
and we thus obtain, in the right hand side, the boundary term we were looking for. Note that the latter is strictly speaking nonlocal as well, because it couples $\rho(0)$ to all values of $a(r^2)$. So, if we require locality, we must set $\rho(0)=0$

Nevertheless, if we decide to keep the boundary term and the $\rho(0)$ parameter, by replacing (\ref{byparts}) in (\ref{boundary_a}) we get a more conventional form of the action
\eqn\label{boundary_c}
\frac{\A}{2\pi Gs}~=~a(b^2)+\bar{a}(0)+2\pi\int_0^{\infty}dr^2\left( -\dot{a}(r^2)\dot{\bar{a}}(r^2)~(\rho(r^2)-\rho(0))-\frac{(1-\dot{\rho})^2}{(2\pi R)^2}\right)\, .
\eqnx
We note at this point that the form (\ref{boundary_c}) of the action yields the same lagrangian equations as in secs. 3 and 4 but with $\rho$ replaced by $\tilde{\rho}\equiv\rho(r^2)-\rho(0)$, which {\it must} vanish at the origin. This is yet another reason for this boundary condition, which can be met by real solutions for $b>b_c$ and by complex ones for $b<b_c$. 

Furthermore, the action is functional of $\tilde{\rho}$ only and, evaluated on the equations of motion takes the form
\eq
\A (b, s)~=~2\pi Gs(a(b^2)+\frac{(I_a+a(b^2)}{2})=2\pi Gs(a(b^2)-I_{\phi})=Gs(2(\chi(L^2)-\chi_b)+1-\frac{1}{t_b})\, ,
\eqx 
in agreement with eq.~(\ref{actionb2}) and with the momentum space relationships.
Thus, in this alternative point of view, $\rho(0)$ is an additive constant in the definition of the field $r^2\dot{\phi}$ in terms of $h\sim (1-\dot{\rho})$, which appears in the effective metric but not in the action. Since this constant  is not there in the perturbative regime $b\gtrsim R$, it should finally be absent altogether.

\end{document}